 \documentclass[authoryear,preprint,12pt]{elsarticle}

\usepackage{float} % added for figures
\usepackage{xcolor}
\usepackage{hyperref}
\usepackage[utf8]{inputenc}
\usepackage{array}
\usepackage{diagbox}
\usepackage{mathtools}
\usepackage{cellspace}
\usepackage{graphicx}

\usepackage{natbib}
\usepackage{mathrsfs} % Provides a \mathscr command
\usepackage{amssymb}
\usepackage{amsmath}
\usepackage{stackengine}
\usepackage{booktabs}
\usepackage{bm}

\journal{journal}
\begin{document}
\begin{frontmatter}

%\title{Control Scenarios for Probabilistic SIR Epidemics on Social-Connection Graphs.
%, and its proxy economic cost assessment. 
%}
%
%\titlerunning{Control scenarios for probabilistic SIR epidemics}
% If the paper title is too long for the running head, you can set
% an abbreviated paper title here
%

%\author{Jan B. Broekaert\inst{1}\orcidID{0000-0002-4039-8514} \and \\
%Davide La Torre\inst{1}\orcidID{0000-0003-2776-0037} }
%% use the tnoteref command within \title for footnotes;
%% use the tnotetext command for theassociated footnote;
%% use the fnref command within \author or \address for footnotes;
%% use the fntext command for theassociated footnote;
%% use the corref command within \author for corresponding author footnotes;
%% use the cortext command for theassociated footnote;
%% use the ead command for the email address,
%% and the form \ead[url] for the home page:
% \title[Competing control scenarios in epidemics on networks.]{Competing control scenarios in probabilistic SIR epidemics on social-contact networks.}
 \title{Competing control scenarios in probabilistic SIR epidemics on social-contact networks.\tnoteref{label1}}
%% \tnotetext[label1]{}
%\author*[1]{\fnm{Jan B.} \sur{Broekaert}}\email{jan.broekaert@skema.edu}
%\author[1]{\fnm{Davide} \sur{La Torre}}\email{davide.latorre@skema.edu}
%\author[1]{\fnm{Faizal} \sur{Hafiz}}\email{faizal.hafiz@skema.edu}
   \author[1]{Jan B. Broekaert} % \corref{cor1} \fnref{label2}
     \ead{jan.broekaert@skema.edu}
   \author[2]{Davide La Torre} % \corref{cor3}\fnref{label4}
     \ead{davide.latorre@skema.edu}
   \author[3]{Faizal Hafiz} % \corref{cor1} \fnref{label2}
     \ead{faizal.hafiz@skema.edu}

\address{Artificial Intelligence Institute, SKEMA Business School, Universit\'{e} C\^{o}te d'Azur, Sophia Antipolis, France}
%\affil[1]{\orgdiv{Institute for Artificial Intelligence}, \orgname{SKEMA Business School}, \orgaddress{\street{60 Fedor Dosto\"ievski}, \city{Sofia Antipolis}, \postcode{06902},  \country{France}}}
  
%% \ead[url]{home page}
% \fntext[label2]{}
%  \cortext[cor1]{Jan B. Broekaert}
%% \affiliation{organization={Artificial Intelligence Institute},
%%             addressline={},
%%             city={},
%%             postcode={},
%%             state={},
%%             country={France}}
%% \fntext[label3]{}

%\institute{Artificial Intelligence Institute, SKEMA Business School, Universit\'{e} C\^{o}te d'Azur, Sophia Antipolis, France\\
%\email{jan.broekaert@skema.edu, davide.latorre@skema.edu}
%}
%
 
% \maketitle              % typeset the header of the contribution
%
\begin{abstract}
%\abstract{
A probabilistic approach to the epidemic evolution on realistic social-contact networks allows for characteristic differences among subjects, including the individual number and structure of social contacts, and the heterogeneity of the infection and recovery rates according to age or medical preconditions.  
Within our probabilistic Susceptible-Infectious-Removed (SIR) model on social-contact networks,  we evaluate the \emph{infection load} or \emph{activation margin} of various control scenarios;  by confinement, by vaccination, and by their combination. 
We compare the epidemic burden for subpopulations which apply competing or co-operative control strategies. The simulation experiments are conducted on randomised social-contact graphs that are designed to exhibit realistic person-person contact characteristics and which follow near \emph{homogeneous} or \emph{block-localised} subpopulation spreading. 
The scalarization method is used for the multi-objective optimization problem in which both the infection load is minimized and the extent to which each subpopulation's control strategy preference ranking is adhered to is maximized. We obtain the compounded payoff matrices for two subpopulations  which impose contrasting control strategies, each according to their proper ranked control strategy preferences. 
The Nash equilibria, according to each subpopulation's compounded objective, and according to their proper ranking intensity, are discussed. Finally, the interaction effects of the control strategies are discussed and related to the type of spreading of the two subpopulations. 
%}
\end{abstract}

%\keywords{SIR dynamics, social-contact graph, confinement, vaccination, competition, co-operation, pay-off matrix}
%% PACS codes here, in the form: \PACS code \sep code

%% MSC codes here, in the form: \MSC code \sep code
%% or \MSC[2008] code \sep code (2000 is the default)

 \begin{keyword}
% %% keywords here, in the form: keyword \sep keyword
 SIR dynamics  \sep social-contact graph \sep confinement \sep vaccination \sep competition \sep co-operation \sep pay-off matrix
% %% PACS codes here, in the form: \PACS code \sep code

% %% MSC codes here, in the form: \MSC code \sep code
% %% or \MSC[2008] code \sep code (2000 is the default)

 \end{keyword}
% \maketitle

\end{frontmatter}
 \section{Introduction and literature review}
 The COVID-19 pandemic has stimulated a novel interest in models of epidemics \citep{BertozziEtAl2020}. Epidemiological models have guided governments and policy makers to develop policies  which include measures of social distancing, lockdowns and vaccination campaigns, all in order to control the risk of an expanding infection. Policies to minimize epidemic burden have in practice required a \emph{modulated} deployment because of regional differences,  behavioral obstacles or social concerns \citep{MacIntyreEtAl2021,Reicher2020}.  Guided by socio-political principles, specific preferences for control policies may differ in subpopulations, and optimizing the epidemic outcome under variable objectives may result in diverse - or competing - strategies in comparison to co-operative strategies. 
In order to assess the epidemic burden when different control strategies are deployed in subpopulations, and when each individual again could exhibit some deviating protective behavior, and could express proper disease related characteristics, a detailed population structure must be implemented in the epidemic diffusion model.   Our present study therefor extends the classical compartmental epidemic model with a  social-contact network structure at the scale of the individual, in order to encompass such individual diversities.\\  
Starting with % Kermack and McKendrick 
\citet{KermackMcKendrick1927}, several models have been elaborated which  subdivide   a population into designated  compartments into which individuals are allocated according to their disease status \citep{Hethcote2000,Choisy2007}. The classical Susceptible-Infected-Removed (SIR) approach is a  basic compartmental epidemic model. In this model individuals are allocated  either to  the \emph{susceptible} (S), the \emph{infected} (I), or the \emph{removed} (R)  meta-population. The extensions of simple epidemic models with structured networks of individual agents (or meta-populations) \citep{Barabasi2013,Newman2018} have successfully been employed in many fields which study phenomena where interrelationships matter \citep{Perc2021} and, include, biological demographic dynamics \citep{Montagnon2019}, international trade \citep{Banerjee2013}, technology diffusion \citep{Eaton1999}, information spreading \citep{LiuZhang_2014,LiEtAl2017}, corruption spreading \citep{kolokoltsovEtAL2017} and contagion in financial markets \citep{DemirisEtAl2012}.
In all these cases the adoption of network models has led to new perspectives and novel insights.
In our present study we develop an extension of the SIR model on social-contact networks with subpopulations adhering to competing epidemic control strategies. 
In a previous work on infectious disease propagation, a probabilistic approach to SIS dynamics on coupled social-contact and economic production networks was explored \citep{BroekaertEtAl2021}.  In particular we applied the \emph{vector logistic equation} on a graph for the infection-recovery dynamics to preserve the probabilistic interpretation of the S- and I-propensity of each individual node (see also \citet{LajmanovichEtAl1976,GerbeauEtAl2014,PrasseEtAl2021}).
The adoption of a network-based approach to modeling an epidemic allows the description of patterns of interaction among individuals - the hetereogeneity of the social-contact network -   or designate subpopulations with specific group properties. In a social-contact network, the links describe all person-person interactions by which the contagious disease can potentially spread. This physical contact network - based on momentarily shared locations \citep{Barrett2009} -  should be distinguished from  the `social network', which typically includes online  besides off-line relationships. Hence in general, a social-contact network spans a very different graph over a population of nodes, both by linkage structures and degrees. \\
The standard SIR-approach models the evolution of an epidemic based on splitting a population  into three different compartments $\{S,I,R\}$. Denoted by $S$ are those individuals who are still untouched by the disease, and who are `susceptible' to infection by a contact  with infected individuals. Compartment $I$ is composed of the  `infective' individuals, who can transmit the disease and who can naturally or therapeutically recover from the disease.  A simple  assumption about infectious contacts sets it proportional to the product of the sizes of the compartments $S$ and $I$ \citep{KermackMcKendrick1927}. 
 Finally, in the `removed' category both those individuals are allocated who have  recovered  and who originate from compartment $I$, and those who have been vaccinated and acquired immunity to the disease.  The classical SIR model for the respective population fractions, $\{s,i,r\}$, reads as
\begin{eqnarray}
\frac{d s}{dt} \ = \ -  \beta  i  s,    \ \ \  \ \ \ \frac{d i}{dt} &=&  \beta i s - \delta i, \ \ \  \ \ \  \frac{d r}{dt} \ = \ \delta i, \label{eq:sir}
\end{eqnarray}
with $\beta$ being the infection rate and $\delta$ the recovery (removal) rate of the disease for the given population. Remark that following the dynamical equation for the $s$ fraction, the latter must be a monotonously decreasing function, and asymptotically in time the $i$ fraction will exponentially reduce to zero. The SIR model hence always entails a recovered  asymptotic outcome to an epidemic outbreak of an infectious disease.

% The SIR dynamics depends on the following number $R_0$ called the basic reproduction number or basic reproduction ratio:

% \begin{equation}
% R_0 = \frac{\beta}{\delta}
% \end{equation}

In our present development of a SIR epidemic on a graph we explore control scenarios that influence the progression of the epidemic at the individual level, rather than e.g. age related  meta-groups of the population as in recent work by \citet{Nakamura2021}. This more detailed dynamic approach can take into account the socio-spatial distribution of individuals and the natural differences of  infection  and recovery capacities of each individual, and also allows to implement associated protective strategies for each subpopulation - or the lack thereof.  Such approaches with a socio-spatial structure of the population by using  complex graph topologies have been successfully developed in various domains  \citep{Keeling2005,Zhou2006,Newman2002,GaneshEtAl2005,LuEtAl2017}.
In particular, our approach aims to describe  the epidemic dynamics immediately at the level of the individual's intrinsic susceptibility, infection and recovery probability, similar to probabilistic  Markov or quantum-like system descriptions, e.g. \citep{BusemeyerBruza2012,WangEtAl2013,Broekaert2020,WangEtAl2003}. The probabilistic interpretation of the  graph state vectors of the nodes  allows the expression of any SIR-related  \emph{expectation} quantity on any targeted subset of nodes. The probabilistic approach then prescribes for some function $f$ over the graph; 
% $\langle \mathbf{f} \rangle_{S,t}$ =  $\sum_{k \in D}
% S_k(t) f_k$, $\langle \mathbf{f} \rangle_{I,t}$ =  $\sum_{k \in D}
% I_k(t) f_k$ and $\langle \mathbf{f} \rangle_{R,t}$ =  $\sum_{k \in D}
% R_k(t) f_k$ 
$\langle \mathbf{f} \rangle_{S,t}$ =  $\sum_{k \in D}
p_{S_k}(t) f_k$, $\langle \mathbf{f} \rangle_{I,t}$ =  $\sum_{k \in D}
p_{I_k}(t) f_k$ and $\langle \mathbf{f} \rangle_{R,t}$ =  $\sum_{k \in D}
p_{R_k}(t) f_k$ 
for the respective SIR-averages at a given time $t$ and targeted node set $D$ of some subpopulation.\\
Recent work has explored the impact of subpopulation differences, e.g. with competing features, on the epidemic evolution. The work of %Guo et al.  
\citet{Guoetal2020}  points out that subpopulations in networks, which follow different rules and have specific interactions between them, will follow  asymmetric approaches - e.g. in the prisoner's dilemma game, and this will lead to the emergence of a pattern of cyclic dominance.   A multi-layer approach by  \citet{peng2021multilayer} allows the inclusion of a layer of \emph{competing opinions} over the network,  which then influence  the contagion linkages in the layer of the physical epidemic. In the work of  \citet{MassaroEtAl2018}  a mobility network approach is implemented with structured populations and with a specifically defined measure of  system-wide \emph{critical functionality}, which includes  the individual infection risk and the disruption to the system’s functionality in terms of the human mobility deterioration.
Our present approach of a probabilistic SIR model on a network differs in the implementation of the subpopulations which deploy competing or co-operative control strategies, and by assessing their respective outcomes through the specifically defined measure of the \emph{infection load}  - or the related \emph{activation margin} - and the adherence to the preferences for specific control policies in each subpopulation. We first develop the probabilistic approach to SIR-dynamics on a network in the next  Section \ref{sec:Prob_SIR}, subsequently we extend this formalism  to model the epidemic burden when subpopulations deploy deviating control strategies, Section \ref{sec:Competing_CS}, and quantitatively assess co-operative versus competing scenarios when subpopulations express different preferences for specific control strategies, Section \ref{sec:Implementation}.

\section{Probabilistic SIR-dynamics on a social-contact graph  \label{sec:Prob_SIR}}

\begin{figure}
	  \begin{minipage}[c]{0.475\textwidth}
         \includegraphics[width=\textwidth]{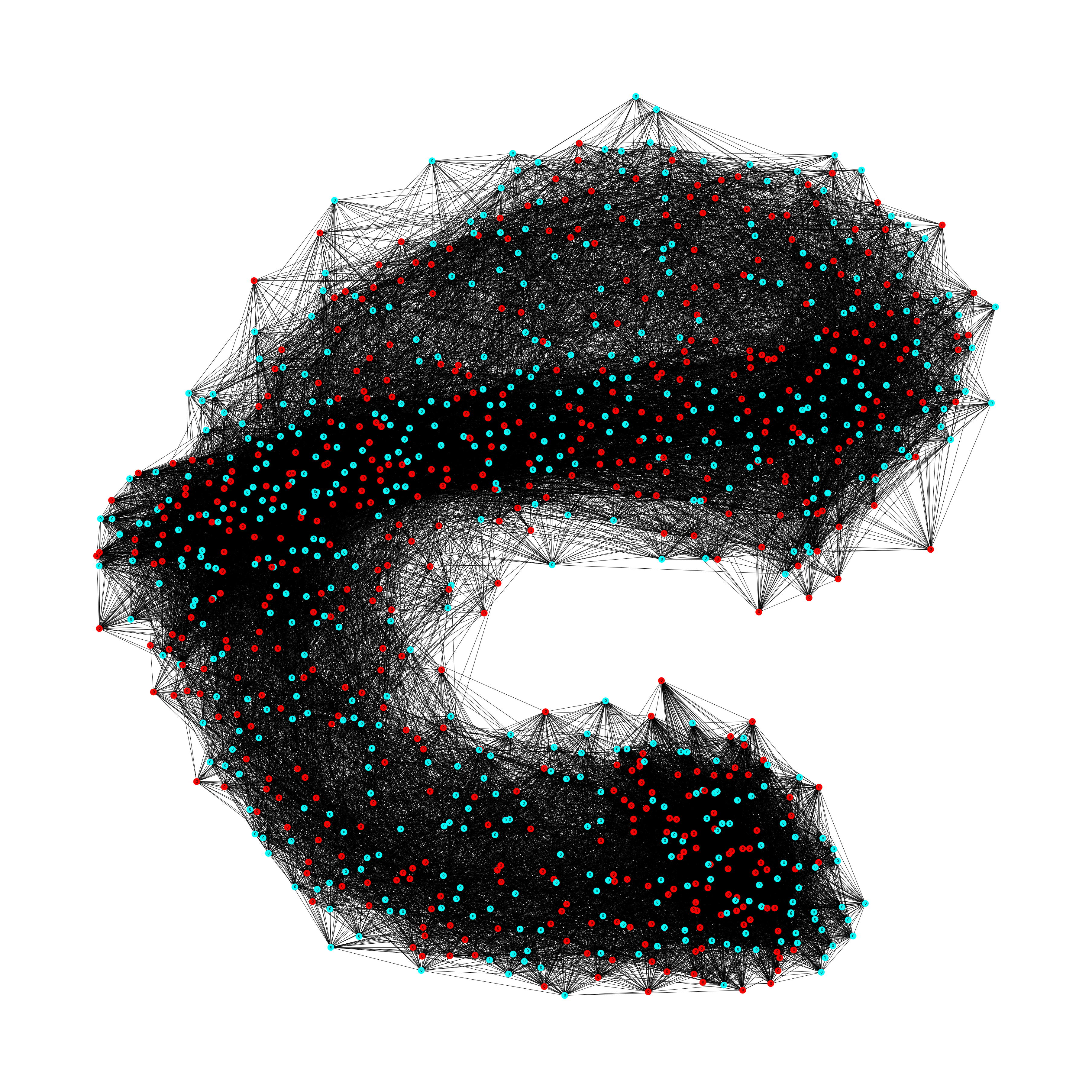}
	    \end{minipage} 
 	     \begin{minipage}[c]{0.05\textwidth}  \hfill	    \end{minipage} 
 	 \begin{minipage}[c]{0.475\textwidth}
	  \includegraphics[width=\textwidth]{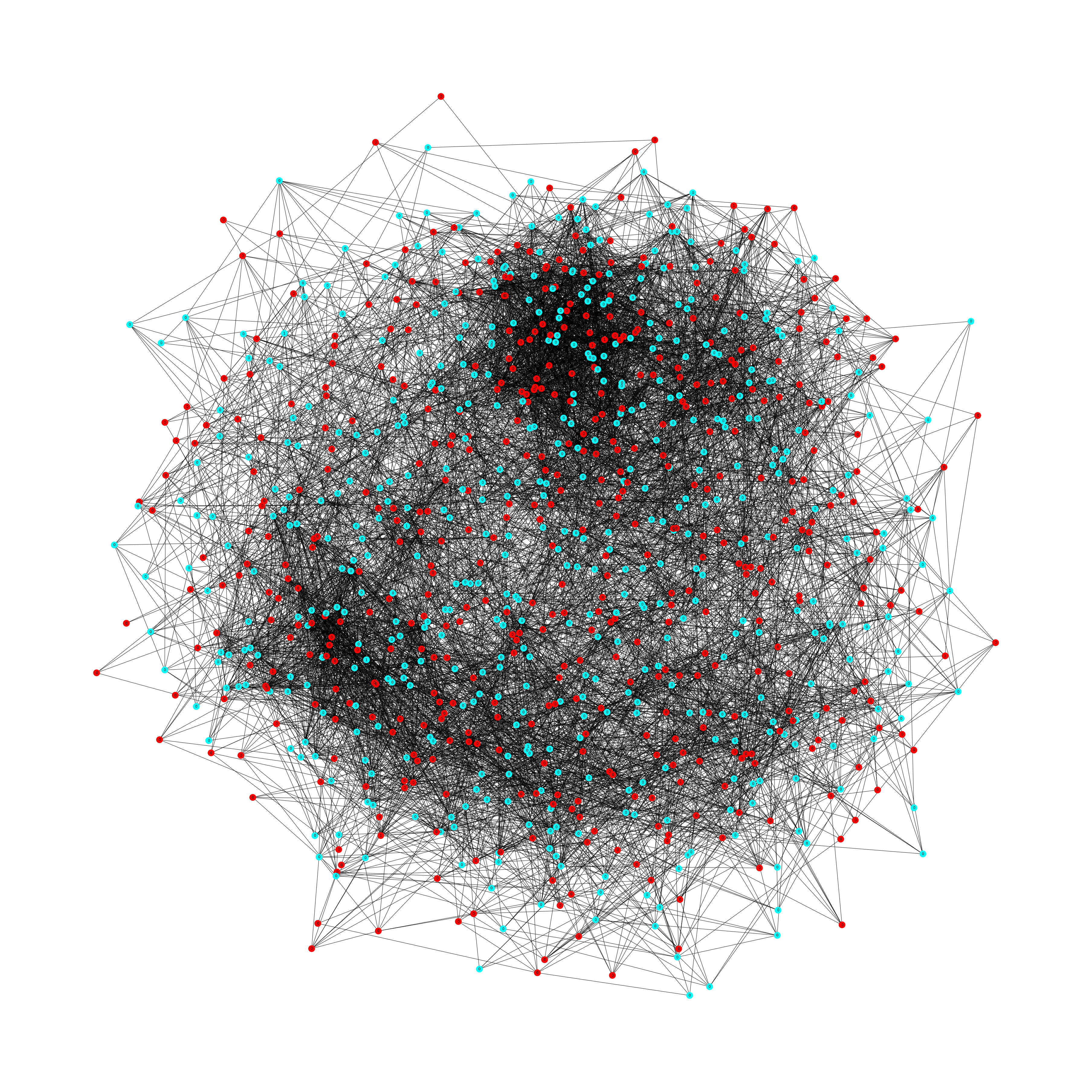}
 	 \end{minipage}
 	 	  \begin{minipage}[c]{0.475\textwidth}
         \includegraphics[width=\textwidth]{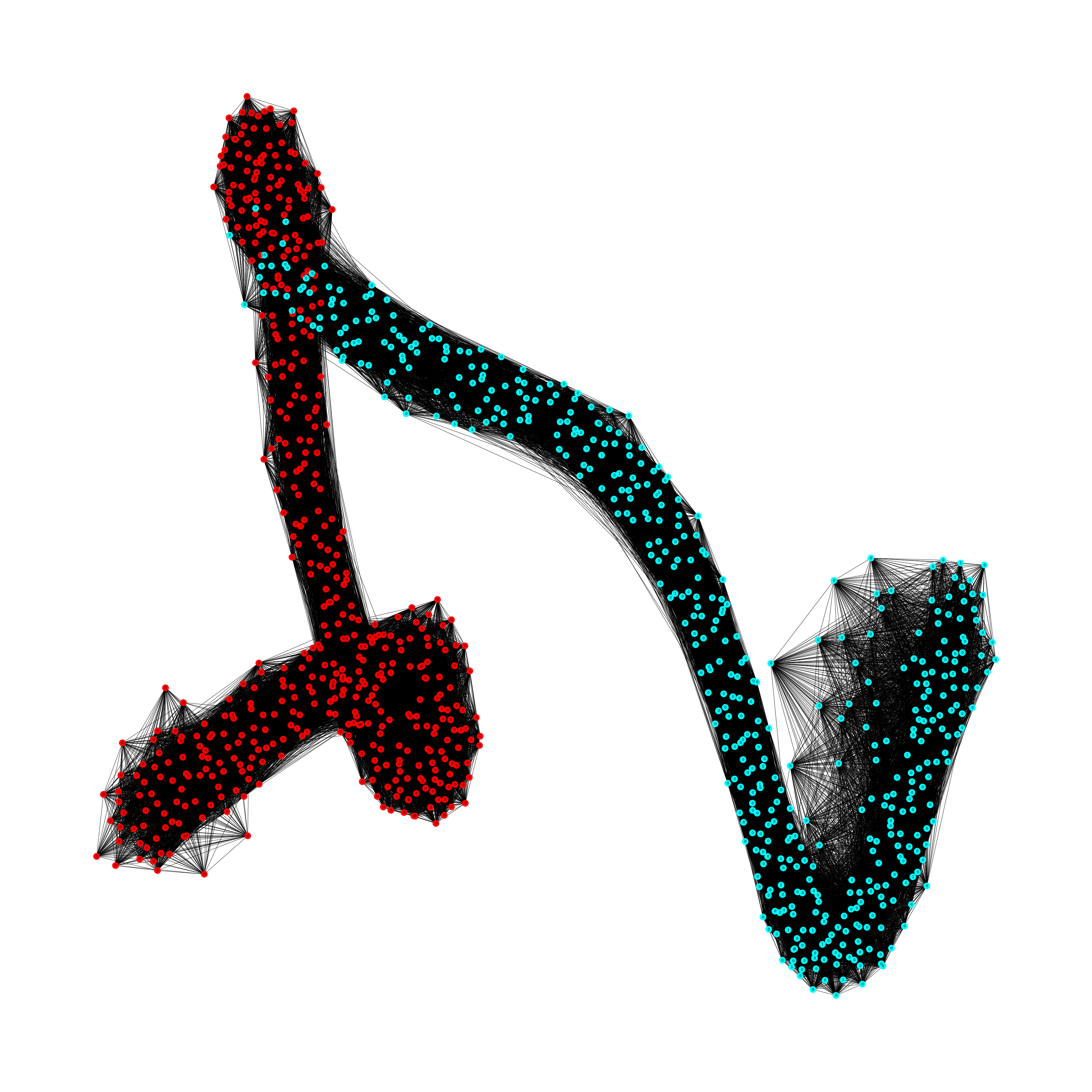}
	    \end{minipage} 
 	     \begin{minipage}[c]{0.05\textwidth}  \hfill	    \end{minipage} 
 	 \begin{minipage}[c]{0.475\textwidth}
	  \includegraphics[width=\textwidth]{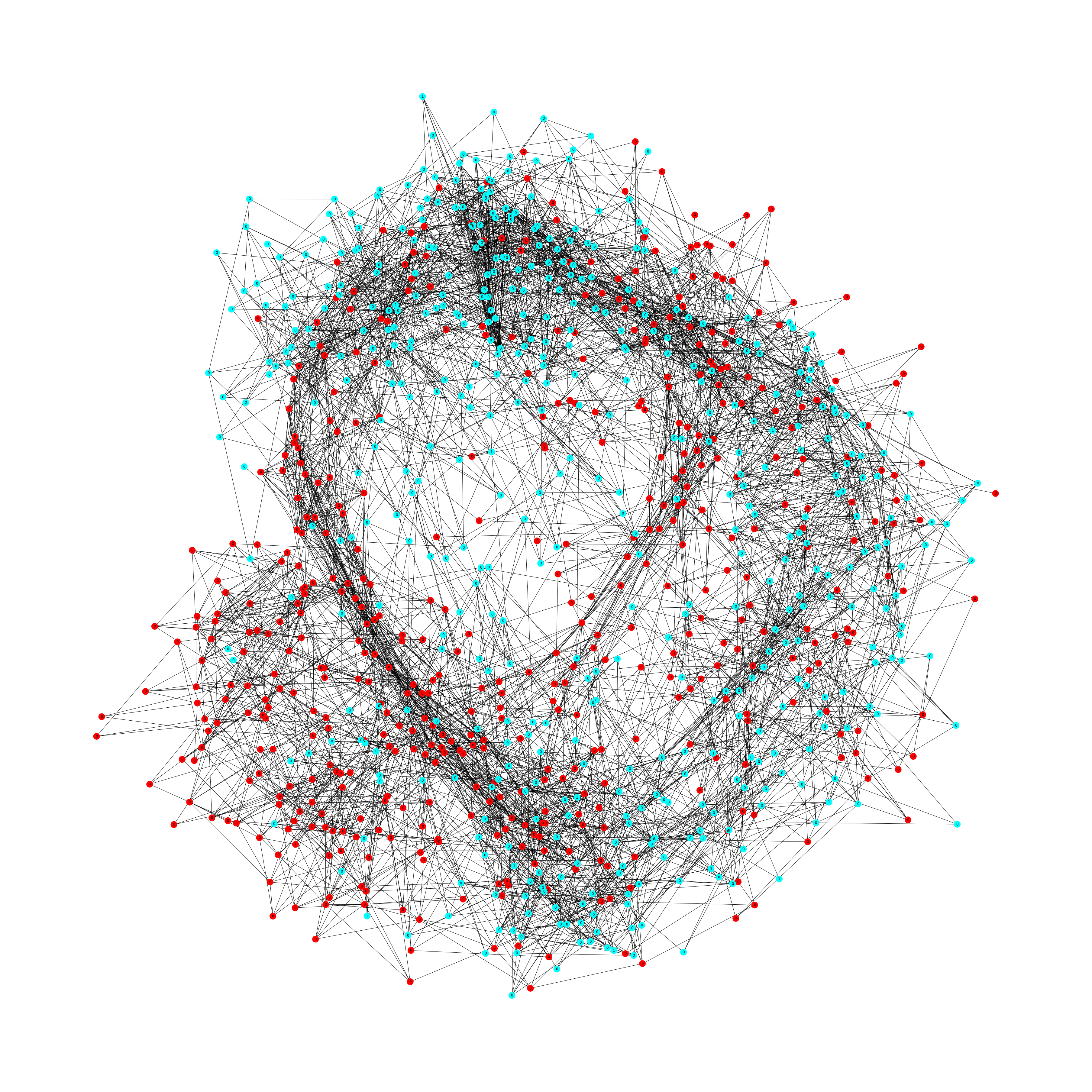}
 	 \end{minipage}
 	 \caption{\scriptsize Examples of two-group (red vs cyan) artificial social-contact graphs (left), $N\!\_{connect}\_A = 50$, following partially randomised homogeneous spreading (top), $skew = .5$,  and block-localised spreading (bottom), $skew = .1$,  and their confinement rendition (right) under co-operating confinement scenario $N\!\_{\text{\it confinement}}\_A = 20$ (top) and  $N\!\_{\text{\it confinement}} = 10$ (bottom). In the unconstrained graphs the mean degree  is 57.76, respectively 96.2. In the confined graphs the mean degree has  respectively decreased to an average of 14.65 and 9.45 remaining contacts. }
%  	 The number of connected social components in the full social graph (adjacency A) is 1 
% The number of connected social components in the confinement social graph (adjacency A_confined) is 5 
% mean degree =  57.76 ,  mean_degree_confined_A =  14.65 ,  N_connect_A =  50    skew =  0.5     mean cluster =  0.26
% Beta function a and b 1.1  and  0.9
% %  	 The number of connected social components in the full social graph (adjacency A) is 1 
% The number of connected social components in the confinement social graph (adjacency A_confined) is 14 
% mean degree =  96.19 ,  mean_degree_confined_A =  9.45 ,  N_connect_A =  50  
% skew =  0.1     mean cluster =  0.59
% Beta function a and b 1.1  and  0.9
 	 \label{Fig:socialgraphs}
\end{figure}

In the probabilistic SIR model on a graph, the individuals in a population are all attributed  with susceptible ($S$), infectious ($I$) and removed ($R$) scalable characteristics. 
The individuals are hence not attributed to one of the three categories by a binary exclusive membership, but reveal a graded association to each characteristic. To this end, to each \emph{node state} three  probabilities, $p_{S_k}, p_{I_k}, p_{R_k}$, are associated  which express the degree to which the given individual node,  $k\in \{1,\cdots, N\}$, pertains to the respective SIR characteristic.
The dynamics of the infected-component, $p_{I_k}$, is determined by the node's individual recovery rate  $\delta_k$, and  by the product of its individual susceptibility, $p_{S_k}$, and all the infectivity propensities $p_{I_l}$ and infection rates  $\beta_l$ of its \emph{neighbouring} nodes, i.e. for which $A_{kl} = 1$, Eq. (\ref{eq:SIR-I}). The adjacency matrix $A$, with $A \in \mathbb{R}^{N,N}, A = A^T $, expresses whether two nodes, $k$ and $l$, have a person-person contact $A_{kl} = 1$, or not $A_{kl} = 0$, and summarizes the basic network architecture of the social-contact graph, see e.g. Fig. \ref{Fig:socialgraphs}.\\
With different subpopulations spread over the network, various masking operations of the adjacency matrix are required to identify \emph{intra-} and \emph{inter-}subpopulation linkages, and to specifically impose associated control strategies per subpopulation (see Sec. \ref{sec:Implementation} and App.).
The individual infection and recovery rate vectors, $\bm{\beta}$ and $\bm{\delta}$, can either be realistically allocated (when such information is available) or randomly attributed following a lognormal distribution. In the latter case a parametrization is required according to a mean infection rate $\beta_{\text{\it avg}}$, a mean recovery rate $\delta_{\text{\it avg}}$ and their respective standard deviations.\footnote{$\beta_{\text{\it avg}}$ corresponds to the $\beta$-value in the SIR model for bulk populations.}
The probabilistic SIR model in terms of individual nodes has the susceptible-component, $p_{S_k}$,  of node $k$ evolve  decreasingly according a $\beta$-weighted product  with its neighbouring nodes' infective components, Eq. (\ref{eq:SIR-S}) (normalised with its degree $d_k$). Additionally this component can evolve by a vaccination intervention, $V_k$.   In our model, the  vaccination of the \emph{randomly} chosen node $k$ at some instant of time, resets the susceptible component to a small value  related to the vaccine's efficiency, Eq. (\ref{eq:vaccination_operator}). 
The \emph{removed}  component, $p_{R_k}$, evolves increasingly according to node $k$'s natural or therapeutic recovery rate   $\delta_k$ from its infectious component, Eq. (\ref{eq:SIR-R}), and when occurring,  by vaccination intervention, $V_k$. The vaccination of a node resets its removed component  mainly to the value of the vaccine's efficiency, Eq. (\ref{eq:vaccination_operator})). The infected-component, $p_{I_k}$,  of node $k$ evolves according the balance of recovery and infection, Eq.(\ref{eq:SIR-I}), and is reset to zero when vaccination occurs, Eq. (\ref{eq:vaccination_operator}). The diffusion of the epidemic over the graph  $\mathscr{G}_A(V,E_A)$ with SIR dynamics and stochastic vaccination is thus given by\footnote{Boldface notation for the {\footnotesize $ N \times 1$} vector $\boldsymbol{\beta}$ and regular notation for the {\footnotesize $ N\times N$} diagonal matrix $\beta$.  }
% \begin{eqnarray} 
% \dot {\mathbf S} &=&   -  {\mathbf V}_{rand.} -    {\mathbf S}    \circ  {\mathbf d}^{-1}   \circ A_t\,     \beta \,  {\mathbf I},     \label{eq:SIR-S}  \\
% \dot {\mathbf I}  &=&   -  \boldsymbol{\delta} \circ {\mathbf I}  +  {\mathbf S}    \circ  {\mathbf d}^{-1}   \circ A_t\,  \beta  \,  {\mathbf I},      \label{eq:SIR-I}  \\
% \dot {\mathbf R} &=&    {\mathbf V}_{rand.}  +  \boldsymbol{\delta} \circ {\mathbf I},   \label{eq:SIR-R} 
% \end{eqnarray}
% \begin{eqnarray} 
% \dot {\mathbf p_S} &=&   -  {\mathbf V}_{rand.} -    {\mathbf p_S}    \circ  {\mathbf d}^{-1}   \circ A_t\,     \beta \,  {\mathbf p_I},     \label{eq:SIR-S}  \\
% \dot {\mathbf p_I}  &=&   -  \boldsymbol{\delta} \circ {\mathbf p_I}  +  {\mathbf p_S}    \circ  {\mathbf d}^{-1}   \circ A_t\,  \beta  \,  {\mathbf p_I},      \label{eq:SIR-I}  \\
% \dot {\mathbf p_R} &=&    {\mathbf V}_{rand.}  +  \boldsymbol{\delta} \circ {\mathbf p_I},   \label{eq:SIR-R} 
% \end{eqnarray}
\begin{eqnarray} 
 {\mathbf {\dot p}_S} &=&   \left\{ -    {\mathbf p_S}    \circ  {\mathbf d}^{-1}   \circ A_t\,     \beta \,  {\mathbf p_I},    {\mathbf V}_{rand.} \right\},    \label{eq:SIR-S}  \\
 {\mathbf {\dot p}_I}  &=&   \left\{ -  \boldsymbol{\delta} \circ {\mathbf p_I}  +  {\mathbf p_S}    \circ  {\mathbf d}^{-1}   \circ A_t\,  \beta  \,  {\mathbf p_I},     {\mathbf V}_{rand.} \right\},    \label{eq:SIR-I}  \\
 {\mathbf {\dot p}_R} &=&   \left\{     \boldsymbol{\delta} \circ {\mathbf p_I},     {\mathbf V}_{rand.} \right\},    \label{eq:SIR-R} 
\end{eqnarray}
where $A_t$ indicates the time-epoch dependent adjacency matrix which can result from the temporary confinement strategies in the network.
Note that $\dot p_{S_j}+\dot p_{I_j}+\dot p_{R_j}= 0$ for all $j$, over the continuous time evolution and the stochastic vaccination operation, Eq. (\ref{eq:vaccination_operator}). This ascertains the conservation of the value of the summed components. With initial conditions $ p_{S_j}(0)+ p_{I_j}(0)+  p_{R_j}(0)= p_{S_{j,0}}+  p_{I_{j,0}}+  p_{R_{j,0}}= 1$, the sum of SIR-components in each node remains equal to 1. One further notices if $p_{I_k} = 1$, then  $p_{S_k} = 0$ and   $\dot  p_{I_k}  =   -   \delta_k  $ shows a decreasing $I$ component. While, if $p_{I_k} = 0$ and if $p_{S_k} \geqslant 0$, then $\dot  p_{I_k}  =      p_{S_k}     {d_k}^{-1}    A_{kj} \beta_j p_{I_j} $ which shows an increasing $I$ component. This ascertains $p_{I_j} (t) \in [0,1]$ at all times. With non-negative SIR-components, the $p_{S_k}$ are monotonically decreasing and $p_{R_k}$ monotonically increasing - except instantaneously at the moment  of the vaccination re-set, see Eqs. (\ref{eq:vaccination_operator}). If  $p_{R_k} = 1$  then $p_{S_k} = 0$, $p_{I_k} = 0$  and   $\dot  p_{R_k} = 0$, hence the $p_{R_k}$ component has reached its maximum value.
%\footnote{The $\mathbf{V}_{rand.}$ vaccination operation will fix the next $R_k$ value to the vaccine's efficiency  weighted value of $(1-e)R_k + e \cdot e $ momentaneously.} 
Similarly, if  $p_{S_k} = 0$ then   $\dot  p_{S_k} = 0 $ and its lowest value is reached. In sum, the components of each  node state  $(p_{S_k}, p_{I_k}, p_{R_k})$  on the social graph satisfy the property of  probabilities, which allows a consistent probabilistic interpretation of the graph-based SIR dynamics with vaccination and confinement, Eqs. (\ref{eq:SIR-S}, \ref{eq:SIR-I},\ref{eq:SIR-R}, \ref{eq:vaccination_operator}), \citep{LajmanovichEtAl1976,GerbeauEtAl2014,PrasseEtAl2021,BroekaertEtAl2021}.\\
The rendition of each individual node's SIR-state over time allows to compare in detail the epidemic burden resulting from various control scenarios. In particular the effectiveness of the control scenarios will be quantified by their resulting \emph{infection load} or \emph{activation margin}, defined as  
\begin{eqnarray}
\mathrm{infection\ load} &=& \frac{1}{N}\sum_j\stackunder{max}{\text{\tiny $t$} } \left(p_{I_j}\right), \label{eq:infection_load} \\
\mathrm{activation\ margin} &=& \frac{1}{N}\sum_j\stackunder{min}{\text{\tiny $t$}}\left(p_{S_j}+p_{R_j}\right), \label{eq:activation_margin}
\end{eqnarray}
Notice that both measures require solving the temporal evolution of SIR-state  for each individual node in the graph, Eqs. (\ref{eq:SIR-S}, \ref{eq:SIR-I}, \ref{eq:SIR-R}). Both measures provide an assessment of the disease impact on each individual over the full time range of the epidemic. This is in contrast with an impact measure of an epidemic outbreak  which would report the maximum value of the average infection probability over time, $\stackunder{max}{\text{\tiny $t$}} \left(\frac{1}{N}\sum_j p_{I_j} \right)$. The infection load reports  a cumulative effect of the epidemic by compounding the possible worst infection impact on each individual. The infection load, Eq. (\ref{eq:infection_load}), hence provides an upper bound to the maximal population infection and gives a measure  of a longitudinal impact of the epidemic.\footnote{An averaged cumulative measure of the infection probability is less informative on the impact of the epidemic since it does not track disease outbreak well due to e.g. individual's peaked infection probabilities. } Similarly, the activation margin  - as defined in Eq. (\ref{eq:activation_margin}) - compounds the lowest `fitness level' of each individual during the epidemic, and hence reduces to the complement of the infection load.

% \begin{figure}
% 	  \begin{minipage}[c]{0.50\textwidth}
%          \includegraphics[width=\textwidth]{Graph_degrees_free.png}
% 	    \end{minipage} 
%  %\begin{minipage}[c]{0.1\textwidth}	    \end{minipage} 
%  	 \begin{minipage}[c]{0.48\textwidth}
% 	  \includegraphics[width=\textwidth]{Graph_degrees_confined.png}
%  	 \end{minipage}
%  	 \caption{\scriptsize  The heavy-tailed degree distribution of the random artificial social-contact graph $\mathscr{G}_A(V,E_A)$ of Fig. \ref{Fig:socialgraphs}, in unrestricted setting (left), and during confinement $\mathscr{G}_{A_{conf}}(V,E_{A_{conf}})$ (right), exhibiting  differences in ego networks and variable compliance with confinement regulation by the individual nodes. }
%  \label{Fig:socialgraphdistributions}
% \end{figure}

\section{Competing and co-operative control scenarios in subpopulations.\label{sec:Competing_CS}}
\begin{figure}
     \includegraphics[width=\textwidth]{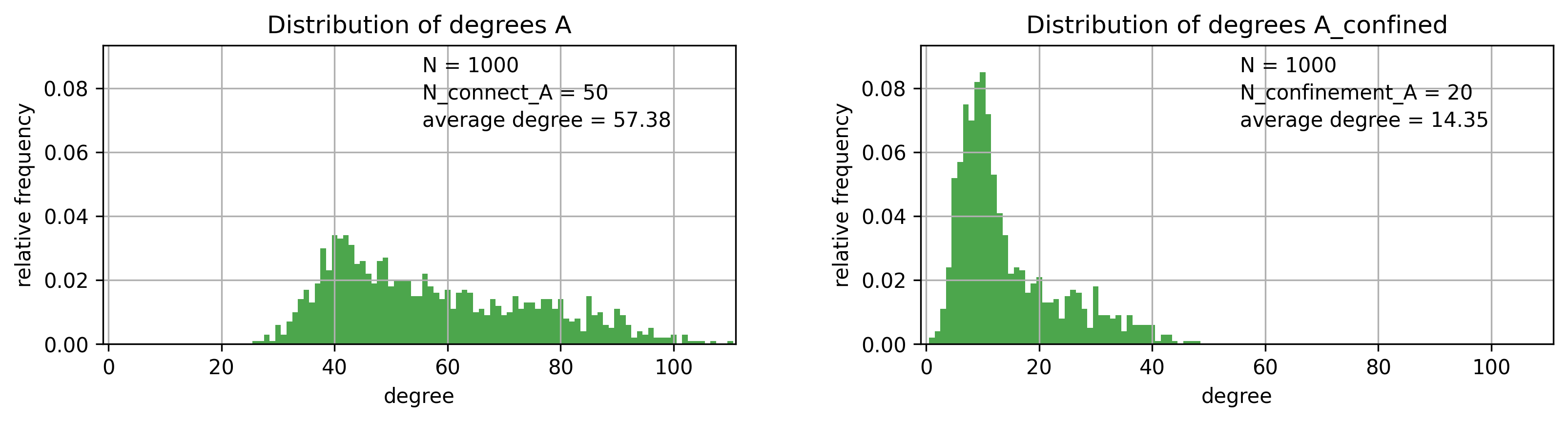}
 	 \caption{\scriptsize  The heavy-tailed degree distribution of the random artificial social-contact graph $\mathscr{G}_A(V,E_A)$ of Fig. \ref{Fig:socialgraphs}, in unrestricted setting (left), and during confinement $\mathscr{G}_{A_{conf}}(V,E_{A_{conf}})$ (right), exhibiting  differences in ego-networks and variable compliance with confinement regulation by the individual nodes. }
 \label{Fig:socialgraphdistributions}
\end{figure}

The epidemic burden of an infectious disease can be partially controlled by deploying policies of social distancing,  lockdowns, curfews and targeted or generalised vaccination campaigns. In order to capture the effect of such policies, epidemic models have included  formalisms, for e.g.   vaccination, shield immunity or quarantine control in a compartmental SEIR-model \citep{XuEtAl2021}, or hierarchical quarantine control \citep{Yang2021}. In our present model we implement both  confinement and  vaccination control strategies and their combination.  Furthermore we allow distinct control strategies for each subpopulation and will consider two variants of the topological spreading of the subpopulations over the social-contact graph.\\
{\bf Confinement.}
In our model we will define the \emph{confinement} approach as any set of measures which results in less contagious transmission of the disease, including e.g.; restricted public hours or perimeters, social distancing, sanitary mask wearing or restricted contact bubbles. The goal of this control measure is to reduce the number of infections and the number of infected individuals through natural and therapeutic recovery, at the end of the confinement time window.
The control strategy by confinement is implemented in the graph model by replacing - during a pre-defined window of time - the adjacency matrix, $A$, parametrized with the threshold connection parameter ($N$\!\_\emph{connect}\_$A$), by the culled adjacency matrix, $A$\_\emph{confinement}, in which the connection parameter has been lowered ($N$\!\_\emph{confinement}\_$A$), see Fig. \ref{Fig:socialgraphdistributions}. In the case of a competing confinement control scenario,  the adjacency matrix receives an adjusted threshold mask for the confined subpopulation only, and only so for the duration of the confinement period (see App.).
\\
{\bf Vaccination.} The control strategy by vaccination is implemented in the dynamics, Eqs (\ref{eq:SIR-S}, \ref{eq:SIR-I}, \ref{eq:SIR-R}), according a stochastic operation, Eq. (\ref{eq:vaccination_operator}), starting at a defined initial time point and running till the full supply of the vaccines is administered.
We make a simplifying assumption, that by its vaccination any node with SIR probability state $(p_S,p_I,p_R)$ will reduce to a weighted linear combination with the optimally vaccinated SIR-state $\left( 1-e,0,e\right)$ at the instance of administration,
\begin{eqnarray}
{\mathbf V}_{\rm rand.}: \qquad \begin{pmatrix} p_S\\ p_I\\ p_R\end{pmatrix}_{\!\!\! t+dt} = e\begin{pmatrix} 1\!-\!e\\ 0\\ e\end{pmatrix} + (1-e)\begin{pmatrix} p_S\\ p_I\\ p_R\end{pmatrix}_{\!\!\! t}, \label{eq:vaccination_operator}
\end{eqnarray}
 where $e$ is the efficiency of the vaccination treatment, $e \in [0,1]$. This operation conserves the unit value of the total probability mass in each node.
A high vaccination efficiency will set the SIR-state pre-dominantly in the \emph{recovered} mode, while a small remaining \emph{susceptible} component allows for possible breakthrough infections to occur. When the vaccination is largely ineffective, $e\approx0$, the SIR-state remains essentially unchanged.
% \begin{eqnarray}
% (S_k, I_k, R_k)_{t+dt}  & = &    e \left( 1-e,0,e \right) + (1-e)(S_k, I_k, R_k)_{t}, \label{eq:vaccination_operator}
% \end{eqnarray}
This operation implies that all types of nodes are vaccinated, even if the probability for $R$ or $I$ is higher than the probability for $S$. The model further assumes a node is  vaccinated only once at most. In practice, after a pre-defined  number ($lag$) of time steps a fixed number ($vacc$\_$batch$) of randomly chosen nodes is vaccinated, the other non-vaccinated nodes and earlier vaccinated nodes continue to evolve according the continous-time SIR dynamics without any vaccination operation. The time lag of administration, the phased batch size and the total amount of vaccines allow to model the vaccination strategy with a realistic slower deployment in comparison to confinement strategies of shorter duration. In the case of competing vaccination control scenarios, the population is masked to only target the concerned subpopulation. When only one of the subpopulations deploys a vaccination control strategy, the full administration capacity is applied to this subpopulation. When more subpopulations choose the vaccination strategy, the administration capacity remains fixed for the whole population.  A joint  vaccination strategy with full availability of vaccines for the whole population hence spreads the fixed administration capacity over all the subpopulations, and thus leads to a smaller number of individuals in each group receiving the vaccination at each lagged time instance, and requires a longer time period to vaccinate the whole population.\\
{\bf Subpopulations topology.}
The structuring of subpopulations in a random contact-graph can be implemented according a number of guiding principles. In our present study we include both structuring based on \emph{homogeneous spreading} - e.g. differences due to political inclination,  and \emph{block localisation}  of subpopulations - e.g. differences due to geographical  concentration. To validly compare the epidemic burden of various control scenarios we  maintained an equal size for the subpopulations.
The configuration with the homogeneous subpopulation spreading was  derived by labeling the nodes along the adjacency diagonal  with consecutive group indices repeatedly.\footnote{e.g. for three subpopulations $012\cdots012$. When the subpopulations are block-localised the spreading is derived from labeling the nodes along the adjacency diagonal repeatedly with the same group index and consecutively  with the next index; e.g. for three subpopulations $0\cdots01\cdots12\cdots2$.}
 A more realistic graph adjacency is subsequently obtained by random partial  shuffling of the node group labels.\footnote{A re-assignment of node group labels is done after sorting the partially randomized node indices - paired with the original group label. The partially randomised node index is obtained by adding a random number  $\sim \mathcal{N}(0,1)$ -  scaled by the \emph{orderliness} parameter set at value 0.1  - to the node index. }\\
{\bf Subpopulation  control policy preferences.}  Socio-political differences in populations (see e.g. \citet{grosEtAl_2021}), or subpopulations, lead to the implementation of different control policies. In our present study we compound the main objective of a decrease of the \emph{infection load}, due to the implemented control strategy, with its proxy cost related to its \emph{preference ranking} and the intensity of this ranking. In the present context we fix the control strategies in an \emph{ad hoc} order
\begin{eqnarray}
&[ \  \text{none},\ \text{confinement},\ \text{vaccination},\ \text{confinement and vaccination} \ ] &
\end{eqnarray}
relative to which, for each subpopulation, the preferred rank of control strategies is denoted.  
In particular, we define two artificial contrasting subpopulations according to their difference in policy preference ranking, e.g. \citet{allcottEtAl2020}:
\begin{eqnarray}
O_1 \ = \ \left[ 1, 2, 3, 4 \right], &\ \ \ \  &O_0 \ = \  \left[ 4, 2, 1, 3 \right]. \label{eq:policyrankings}
\end{eqnarray}
Out of  a continuum of possible policy rankings, we selected these two profiles which evoke a clear degree of partisanship or polarisation,
and of which variations do occur in  real-world socio-political situations \citep{DeaneEtAl_2021,Schaeffer_2021}. The policy ranking profile, $O_0$, of  the `0-group',  \emph{vacc.} $\succ$  \emph{conf.} $\succ$ \emph{conf.$\_$vacc.} $\succ$  \emph{none}, expresses a preference for medical intervention prior to a social effort to diminish transmission and which favors the deployment of a control policy. In contrast, the profile $O_1$ of the `1-group',  \emph{none} $\succ$ \emph{conf.} $\succ$  \emph{vacc.} $\succ$ \emph{conf.$\_$vacc.}, expresses a preference for a `hands-off' approach, with the deployment of a medical intervention strategy only in last resort.  A measure of nuance to the policy preferences is provided by retaining an `intensity', $I$, for the provided ranking.\footnote{Instead of an aggregated preference ranking by a subpopulation, the policy maker's ranking is considered.}
A quantification method for the utility of ranked criteria has been developed in the context of `multi-decision maker' theory, e.g. by \citet{zhangEtAl2004} and \citet{herreraEtAl2001}, and extends the models of \citet{luceEtAl1965}. These models allow the transformation of a preference ranking into multiplicative weights that can be compounded with - in our present case - the infection load, given the intensities, $I_0$ or $I_1$, of these rankings.
Using Saaty's 1-to-9 magnitude scale, with `9' indicating absolute preference of one over another alternative, and `1' indicating the indifference of preference between the two alternatives, the deployment of the competitive control scenarios in the two subpopulations can be assessed at different levels of socio-political primacy \citep{saaty1978}. 
The multiplicative weighting scheme for the four ($n=4$) alternative policies, with intensity $I$, follows 
\begin{eqnarray}
w(i)  &=& \frac{\Pi_{j=1}^n \tau_{ij}^{1/(n-1)}}{\sum_{i=1}^n  \Pi_{j=1}^n  \tau_{ij}^{1/(n-1)}},  \label{eq:policyweights}
\end{eqnarray}
with  $\tau_{ij} = I^{\Delta_{ij}}$ and $\Delta_{ij} = (O_1(i) - O_0(j))/(n-1)$, \citep{zhangEtAl2004}. The weight vectors, $w_0(i)$ and $w_1(i)$, of the policy rankings, Eqs. \ref{eq:policyrankings}, make up the elements of the policy-based component of the compounded payoff matrix, subsection \ref{subsec:CS_compounded_objective}.\\

With differing epidemic progression in the subpopulations under their respective control strategies, we can now gain insight in how these deviating control scenarios impact the proper group and how these strategies cross-interact.  
 We first implement co-operative control scenarios - i.e. in which all nodes are subject to the same sanitary control strategy - and assess the epidemic burden according to the \emph{infection load}.  Subsequently we will assess the epidemic infection load for subpopulations deploying deviating control scenarios, subsection \ref{subsec:Competing_CS}. With a complete infection-load based pay-off matrix in place, thereafter the effect of compounding the objective with differing policy preferences can be assessed, subsection \ref{subsec:CS_compounded_objective}. To efficiently compare the effect of control scenarios we will implement a competitive model with two subpopulations of equal size.

\section{Implementation and simulation results \label{sec:Implementation}}
The network architecture plays a major role in the diffusion and amplitude of an epidemic \citep{GaneshEtAl2005}.
 The social-contact network in our present study was built as a partially sorted random-based social-contact graph \citep{BroekaertEtAl2021}. The implementation with partial sorting of the initial random probability of a connection between two nodes allows more realistic aspects of true person-person networks, similar to those have been reconstructed by e.g. Barrett et al. \citep{Barrett2009}. The resulting adjacency matrix, $A$, of the  graph enhances the number of cliques and clustering in the graph and produces a heavy-tailed degree distribution, Fig. \ref{Fig:socialgraphdistributions}. The graph is parametrized  by the connection number ($N\!\_connect\_A$) to tweak the number of contacts, and the `order variable' ($skew$) which weights the degree of randomness of adjacencies in the graph. An effective connection between two nodes is  established if the connection probability is equal or larger than a given threshold, which depends on the confinement parameters imposed by each subpopulation (details appear in the Appendix). Finally, a weighting procedure on sorted and unsorted connection probabilities with a $skew$-parameter allows to adjust for clustering degree in the graph, App. Eq. (\ref{eq:skew}).\\
 In our simulations experiments we fixed the population size ($N=1000$), and the unconstrained connectivity ($N\!\_connect\_A = 50$). When subpopulations follow competing control strategies partial maskings of the adjacency matrix for \emph{intra} and \emph{inter} subpopulation linkages are used to reconfigure the temporary adjacencies in the dynamics, Eqs. (\ref{eq:SIR-S} - \ref{eq:SIR-R}). Individuals belonging to the same subpopulation  connect according to their associated confinement strategy. For the \emph{inter} subpopulation linkages a \emph{negotiated} connection parameter is used, which equals the average of both subpopulation's connection parameter, App. Eq. (\ref{eq:mixed_threshold}).
For each random social graph in the simulations Experiment 1 and Experiment 2 - e.g. Fig. \ref{Fig:socialgraphs}  and its degree distribution, Fig. \ref{Fig:socialgraphdistributions} - fixed epidemic parameters have been used. In particular, we implemented an epidemic with random lognormal distributed \emph{individual} infection rates  $\beta_{\text{\it avg}} = .08$ ($\beta_{\rm SD}$ = .05) and recovery rates  $\delta_{\text{\it avg}} = .2$ ($\delta_{\rm SD}$ = .05), and with  50 infected nodes ($init\_\text{\it infected}$) at start.\footnote{Remark that the consistent interpretation of the infection rate $\beta$ in the graph model, Eqs. (\ref{eq:SIR-S},\ref{eq:SIR-I},\ref{eq:SIR-R}), and in the bulk model, Eqs. (\ref{eq:sir}), requires re-scaling by a factor $\lambda= \sum_{ij} A_{i,j} / N^2$ according $\beta_{bulk} = \lambda \beta_{graph} $.}
Hence, the subpopulations have not been differentiated with respect to the epidemic parameters but solely  according to their deployed control strategy.
\begin{figure}[h]
         \includegraphics[width=\textwidth]{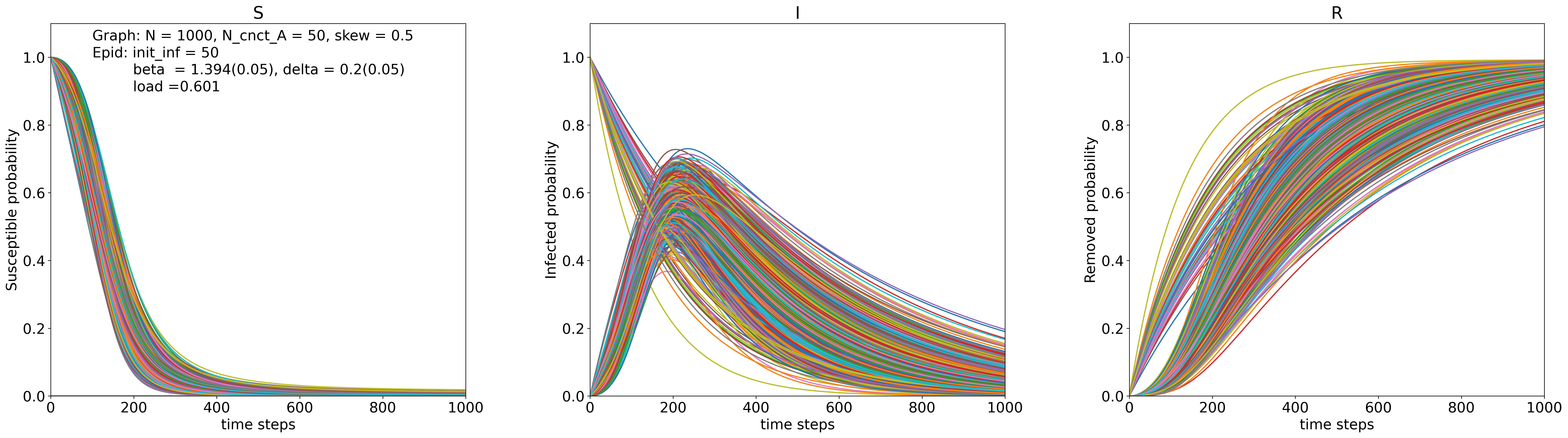}
         \includegraphics[width=\textwidth]{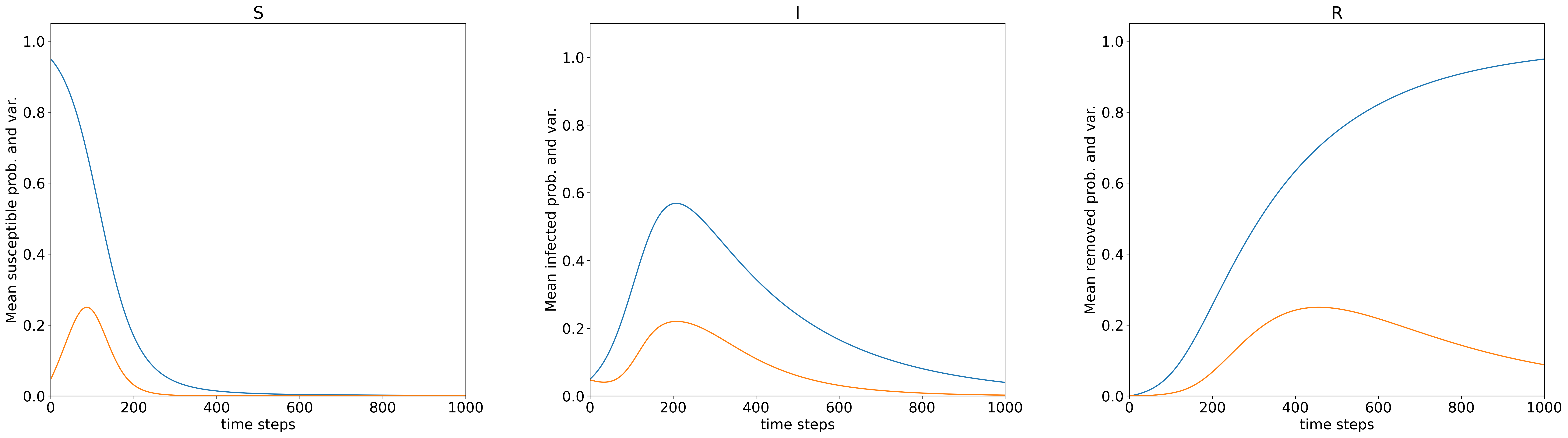}
 	 \caption{\scriptsize {\bf Uncontrolled SIR epidemic}. An illustrative SIR-epidemic evolution on a random social-contact graph $\mathcal{G}(V,E_A)$ with 1000 nodes ($N$), 50 initially infected nodes (\emph{init}$\_$\emph{infected}), and an average social connectivity parameter set at 50 (\emph{N}$\!\_\emph{connect}$\_\emph{A}) causing a .601 infection load. The individual connectivity degree, infection and recovery rates allow a fine-grained analysis of SIR-state progression (top panels) in comparison to its mean-field and variance rendition (bottom panels). }
 \label{Fig:SIR_epid}
\end{figure}
We can then assess the effect of the confinement strategy, vaccination strategy, and combined vaccination-confinement strategy,  by comparing their resulting \emph{infection load},  Eq. (\ref{eq:infection_load}), with its base-value in the no-control strategy. \\
The vaccination strategy   will take   a random selection of nodes from the vaccination start time and will reset this fixed number  of nodes ($vacc$\_$batch$) to the vaccinated state, Eq. (\ref{eq:vaccination_operator}),and repeat this each fixed number of time steps ($lag$) till the vaccine supply ($vacc\_tot$) is finished.\\
For nodes on which no vaccination operation is completed at a given time instance, the SIR dynamics, Eqs. (\ref{eq:SIR-S}, \ref{eq:SIR-I}, \ref{eq:SIR-R}), reduces to a system of first-order vector differential equations which allow an iterative numerical solution.
Following an initialisation of the SIR-states $\mathbf{p_S} (0) = \mathbf{p_S}_0$, $\mathbf{p_I} (0) = \mathbf{p_I}_0$ and $\mathbf{p_R} (0) = \mathbf{p_R}_0$ an incrementally updated solution is  obtained through the operation of an iteratively updated propagator;\footnote{The required size of the time step $dt$ is controlled with respect to the tolerance on the deviation of the probability conservation in each node. With $dt= 0.02$, $steps  = 1000$,   the present graph model sets a tolerance of $\Delta p_{tot} \approx 0.03$.}
\begin{eqnarray}
\mathbf{p_S} (t+dt)  & = & \mathbf{p_S} (t) - dt \, \mathbf{p_S}(t) \circ \mathbf{d}^{-1} \circ  A(t) \beta \, \mathbf{p_I}(t) \\
\mathbf{p_I} (t+dt)    & = &  (\mathbf{1}-dt \,\bm{\delta}) \circ \mathbf{p_I}(t) +  dt \, \mathbf{p_S}(t) \circ \mathbf{d}^{-1} \circ  A(t) \beta \, \mathbf{p_I}(t)  \\
\mathbf{p_R} (t+dt)   & = & \mathbf{p_R}(t) + dt \,\bm{\delta} \circ \mathbf{p_I}(t)  
\end{eqnarray}
In order to compare the effectiveness of the vaccination control and the confinement control and their combination,  we assess the epidemic burden for \emph{co-operative}  scenarios in  simulation Experiment 1, (Sub-Sec. \ref{subsec:Co-operative_CS}).  The configurations with subpopulations that deploy \emph{competing} scenarios are assessed in Experiment 2, (Sub-Sec. \ref{subsec:Competing_CS}). All scenarios are finally assessed according to the compounded objective of the infection load reduction and a measure of adherence to the preference ranking of the control strategy, (Sub-Sec. \ref{subsec:CS_compounded_objective})

\begin{figure}[H] % changed h to H
         \includegraphics[width=\textwidth]{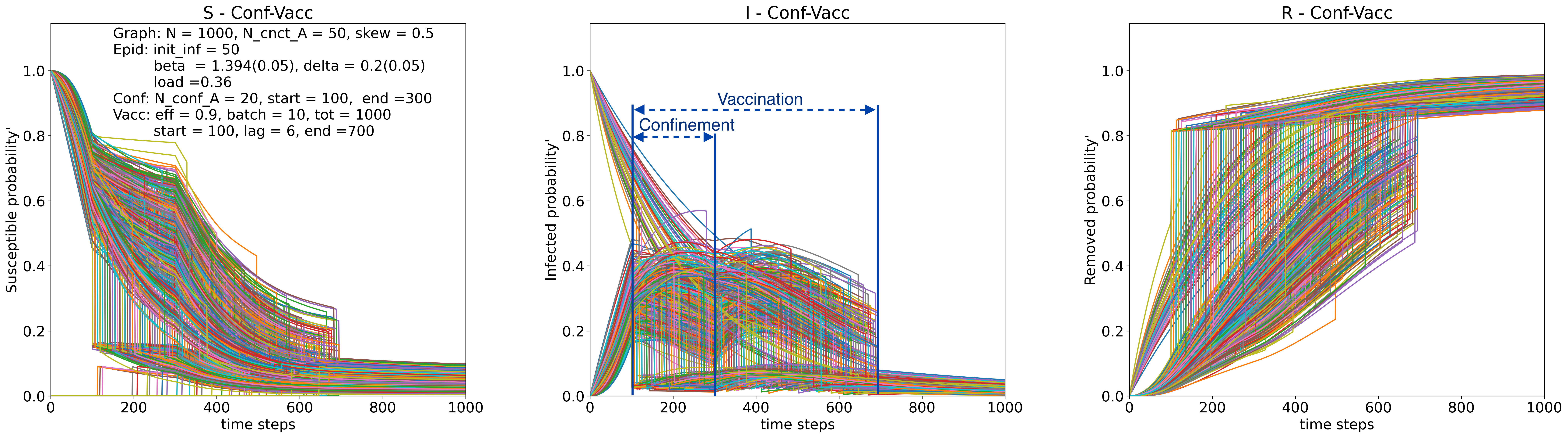}
         \includegraphics[width=\textwidth]{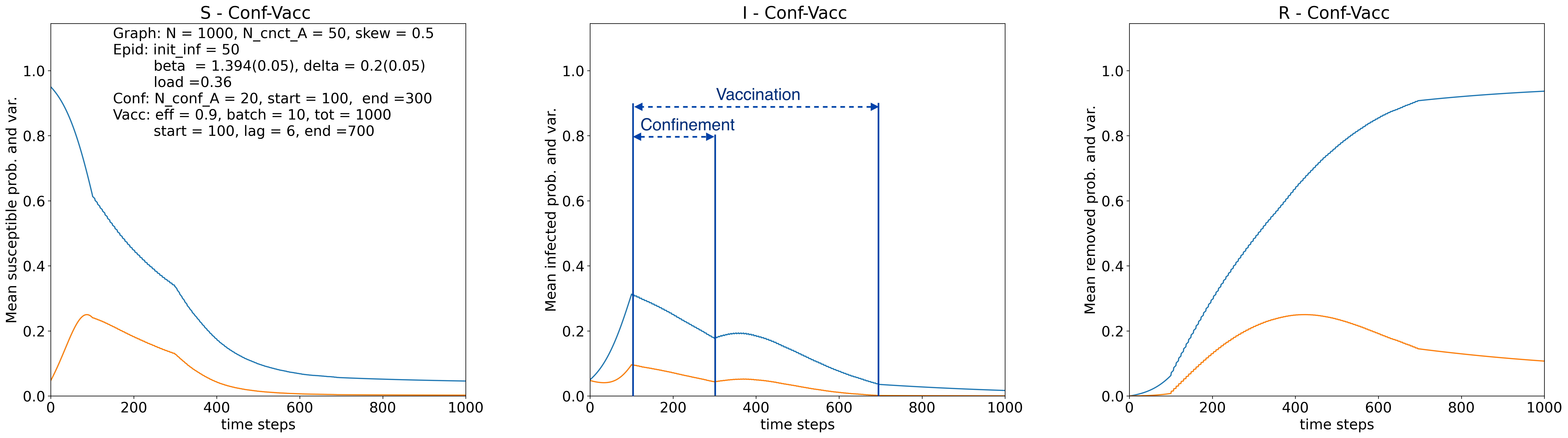}
 	 \caption{\scriptsize {\bf Co-operative controlled SIR epidemic}. The relative optimal control strategy of `early' combination of vaccination and confinement on the SIR-epidemic, Fig. \ref{Fig:SIR_epid}, for the random social-contact graph $\mathcal{G}_A(V,E_A)$, Fig. \ref{Fig:socialgraphs}, results in an infection load of .36, reduced from .601 without control strategy, Tab. \ref{tab:infection_load}.  SIR-state progression (top panels) in comparison to its mean-field and variance rendition (bottom panels). Vaccination and confinement start at time step 100. The confinement reduces the contact parameter from 50 to 20 for a duration of 200 steps. The vaccination randomly selects 10 nodes every 6 time steps and ends after 600 steps with the depletion of the 1000 available doses.  }
 \label{Fig:SIR_epid_Vac_conf}
\end{figure}

\subsection{Co-operative control scenarios for the single objective of infection load reduction \label{subsec:Co-operative_CS} }
 For Experiment 1, we note that with the fixed configuration of the social graph type and epidemic, Fig. (\ref{Fig:socialgraphs}),  a maximal graph-averaged infection is reached for $t \approx 200$ when `no control' is deployed, Fig. (\ref{Fig:SIR_epid}).  With respect to this observation, our study proposes three confinement strategies that imply the same proxy cost as implied by the duration (\emph{steps\_{conf}\_duration}), and the severity of the confinement, (\emph{N\!\_confinement\_A}) but only differ by their starting point in time; `Early', `Mid' and `Late'  (\emph{steps\_conf\_start} = 100,  200, 300 respectively).
\begin{table}[htbp]
   \centering

   \begin{tabular}{@{} lcccc @{}} %  
      \toprule
 
              & No              & Conf          & Vacc              &  Conf--Vacc\\
      \cmidrule(lr){2-2}   \cmidrule(lr){3-3} \cmidrule(lr){4-4} \cmidrule(lr){5-5}
     `Early'   & 0.603 (0.002)  & 0.444 (0.005)   &  0.566  (0.003) &  {\bf 0.365 (0.004)} \\ 
     `Mid'     & 0.603 (0.002)  & 0.596  (0.003)  &  0.602 (0.002)  & 0.596  (0.003)\\ 
     `Late'    & 0.603 (0.003)  & 0.603 (0.003)   &  0.603 (0.003)   & 0.603 (0.003)\\
      \bottomrule
   \end{tabular}
%   EARLY
%[[array([0.60308068, 0.44410301, 0.56624438, 0.36467461])],
% [array([0.00226922, 0.00459702, 0.00325323, 0.00382509])]]
%   MID
%[[array([0.60317716, 0.5963375 , 0.6024172 , 0.59633278])],
% [array([0.00235256, 0.00267177, 0.00233504, 0.00267188])]]
% LATE
%[[array([0.60311181, 0.60311179, 0.6031118 , 0.60311179])],
% [array([0.00259407, 0.00259402, 0.00259405, 0.00259402])]]
  \caption{The averaged {\bf infection load} and standard deviation, Eq. (\ref{eq:infection_load}), in nine basic control strategies uniformly applied on  random social-contact graphs $\mathcal{G}_A(V,E_A)$, of type Fig. \ref{Fig:socialgraphs}, with  SIR-epidemic, of type Fig. \ref{Fig:SIR_epid}.  
   One instance of the SIR pandemic evolution of the relative optimal strategy `early' combination of confinement and vaccination with infection load .3791 is given in Fig. \ref{Fig:SIR_epid_Vac_conf}. Reported infection loads are averages of 50 repetitions of random social-contact graphs and epidemic with identical parametrization $N=1000$, $N\!\_connect\_A=50, skew = .5$, $\beta_{\text{\it avg}} = .08$ ($\beta_{\rm SD}$ = .05), $\delta_{\text{\it avg}} = .2$ ($\delta_{\rm SD} =.05$) and $vacc\_tot$ = 1000.}
   \label{tab:infection_load}
\end{table}
 
% \begin{table}[htbp]
%   \centering
%   \begin{tabular}{@{} lcccc @{}} %  
%       \toprule
 
%               & none & confinement  & vaccination &  confinement-vaccination\\
%       \cmidrule(lr){2-2}   \cmidrule(lr){3-3} \cmidrule(lr){4-4} \cmidrule(lr){5-5}
%      `Early'   &  0.391  (0.003)  &  0.552 ( 0.004)   &  0.428 (0.003)&  {\bf  0.631 (0.003)} \\ 
%      `Mid'     &  0.390  (0.002 )  &  0.397 (0.003)   &  0.391 (0.002) &  0.397 (0.003)\\ 
%      `Late'    &  0.390  (0.003)  &   0.390  (0.003)  &  0.390  (0.003) &  0.390  (0.003)\\
%       \bottomrule
%   \end{tabular}
% %   EARLY
% % [[array([0.39053649, 0.55175577, 0.42761855, 0.63058173])],
% % [array([0.00271433, 0.00437316, 0.00315538, 0.00349403])]]
% %  MID
% %[[array([0.39021724, 0.3970376 , 0.39097355, 0.39704226])],
%  %[array([0.00242176, 0.0027332 , 0.00240441, 0.00273326])]]
% % LATE
% %[[array([0.39029109, 0.39029111, 0.3902911 , 0.39029111])],
% % [array([0.00265628, 0.00265623, 0.00265626, 0.00265623])]]
%   \caption{Averaged {\bf activation margin} and standard deviation, Eq. (\ref{eq:activation_margin}), in nine basic co-operative control scenarios for random social-contact graphs $\mathcal{G}_A(V,E_A)$. The network and epidemic parameters correspond to Tab. \ref{tab:infection_load}.}
%   \label{tab:activation_margin}
% \end{table}
For the vaccination strategies we have explored three deployments that again imply the same proxy cost as implied by the total number of  1000 vaccines, (\emph{vacc\_tot}), and a number of 10 administered vaccines  (\emph{vacc\_batch}) at the start of each next lag of 6 time steps (\emph{steps\_vacc\_lag}). Also, the effectiveness of the vaccine, $e = .9$, is kept fixed, and again the only difference is the starting point of the vaccination campaign in time, `Early', `Mid' or `Late'. With the given parameters, the vaccination campaign can  cover the full  population and requires 600 time steps (\emph{steps\_vacc\_duration}) to complete. A configuration with a limited  availability of doses can be derived from the competing control scenarios in Experiment 2.    \\
For the combined  vaccination-confinement strategies we have - to simplify comparison - superposed the corresponding separate vaccination and confinement strategies. 
The simulation experiment ($n_{\rm iter} = 50$) reveals that the basic control strategies show a strong dependence of the epidemic  burden on the starting time point of intervention with respect to a presupposed peak of the epidemic, see Tab  \ref{tab:infection_load}. The implemented confinement strategy and vaccination strategy show a clear gain in the infection load for the former strategy, -.159 vs -.037. If no interaction of the strategies occurs the combined strategy is expected to deliver a gain equal to the sum of gains of the two separate strategies. The observed gain from the combined confinement-vaccination strategy at -.238 shows a significant non-additive interaction of the two control components to the size of -0.042. % (-.238-(-.159-.037)).\\
% The observed activation margins, Tab. \ref{tab:activation_margin}, reveal a very similar response to the control strategies and will hence not be reported in further experiments here.\\
Given that the control strategies would come at the same proxy cost and the observation of an  increasing infection load with starting time, Tab. \ref{tab:infection_load}, the guideline is to intervene as early as possible with the epidemic control policy, and to prefer  the synergy of a combined confinement-vaccination strategy above a plain confinement strategy which proves more effective than the plain vaccination strategy in this implemented type of SIR-epidemic. 

\subsection{Competing control scenarios for the single objective of infection load reduction \label{subsec:Competing_CS} }
The epidemic burden is highly dependent on the parameters of a control strategy, it is therefore crucial to insert the specific constraints of the strategies that are in place in the different subpopulations of the network. The effect and cross-interaction  of deviating scenarios will moreover depend on the topological spreading of the subpopulations over the random contact-graph. In our present study we include both; configurations based on a homogeneous spreading and configurations based on a block localisation of the subpopulations.  These two subpopulation configurations are adapted to respectively model individual differences that can often be considered independent of geographical location within a community -  like political inclination or age group, versus individual differences that do depend on location like regional constituency or national identities.  The effect of deviating control scenarios is now assessed by applying subpopulation masks to the population in order to obtain group-specific infection loads. \\ %(and activation margins).\\
%To obtain more realistic population spreading, each node's  group label is assigned by random partial-shuffling of homogeneous and blocked distributions over the  diagonal of the adjacency matrix $A$.\\
In simulation Experiment 2 all the possible competitive and co-operative strategies of confinement, vaccination, confinement-vaccination or no-control are  assessed for two subpopulations.  Given the effectiveness of the tested control strategies in Experiment 1, Tab. \ref{tab:infection_load}, the `Early' time setting is retained. Further, we let the vaccine be available for the full population ($vacc\_tot$ = 1000), since a no-vaccination strategy can be adhered to by a subpopulation.

\begin{table}[H] % changed h! to H
\centering
{
\renewcommand\arraystretch{1}

\begin{tabular}{|c*{4}{|c}|}
\multicolumn{3}{c}{} \\
\hline
& No$_1$ & Conf$_1$ & Vacc$_1$ & Conf-Vacc$_1$\\
\hline
No$_0$  & 0 & \backslashbox[]{-.063}{-.135} &\backslashbox[]{-.020}{-.054}&\backslashbox[]{-.078}{-.191}\\ 
\hline
Conf$_0$ & \diagbox[]{-.135}{-.063} & {-.160$^\star$} & \diagbox[]{-.179}{-.121} &\diagbox[]{-.220}{-.254} \\
\hline
Vacc$_0$ & \diagbox[]{-.054}{-.020}  & \diagbox[]{-.121}{-.179}     &  -.037$^\star$    &  \diagbox[]{-.100}{-.184}\\
\hline
Conf-Vacc$_0$ & \diagbox[]{-.191}{-.078}   &     \diagbox[]{-.254}{ -.220}           &   \diagbox[]{-.184}{-.100}    &   {\bf {-.238$^\star$}}\\
\hline
\end{tabular}
}
\vskip 3mm
\caption{The pay-off matrix of the averaged {\bf infection load} change - with respect to the infection load .602 of the No$_0$-No$_1$ co-operative scenario - for competing two-group control scenarios with partially shuffled-{\bf homogeneous} group spreading in random social-contact graphs $\mathcal{G}_A(V,E_A)$, ($n_{\rm iter} = 50$, SD $\approx$ .003 - .007). The network and epidemic parameters correspond to Tab. \ref{tab:infection_load}, `Early' configuration.  The co-operative control scenarios ($^\star$) result in an indistinguishable epidemic burden under the homogeneous spreading of the subpopulations.}
  \label{tab:homogenous_groups_infection_loads}
\end{table}
The major observations for the control scenarios with the \emph{homogeneous} spreading of the two subpopulations are given by the pay-off matrix of the group-specific infection load change, with respect to the no-control configuration, Tab. \ref{tab:homogenous_groups_infection_loads}.
% Nash : If each player has chosen a strategy  and no player can increase their own expected payoff by changing their strategy while the other players keep theirs unchanged, then the current set of strategy choices constitutes a Nash equilibrium.
First; the optimal control solution  and  the Nash equilibrium in the minimisation of the epidemic burden by the two subpopulations is to deploy a co-operative   Confinement-Vaccination strategy, Conf-Vacc$_0$-Conf-Vacc$_1$, resulting in a -.238 decrease of the infection load for both subpopulations. Neither subpopulation can improve its proper infection load without increasing the other subpopulation's infection load. Other co-operative strategies are less opportune, conform to the observations in Experiment 1. Second; a strong between-group interaction is apparent in the epidemic load. When one subpopulation changes strategy this affects the epidemic burden not only for the \emph{in-group} but also for the \emph{out-group}. Because of the abundant between-group adjacencies,  the deviating control scenarios frequently inject relatively exogenous SIR-levels in each subpopulation.
%--- In the competing control scenarios No$_0$\_Conf$_1$,  No$_0$\_Vacc$_1$ and No$_0$\_Conf-Vacc$_1$, 
This is most visible in the competing scenario when one subpopulation holds on to the no-control strategy and it still benefits from an epidemic burden improvement  of -.063 in  No$_0$\_Conf$_1$, -.020 in No$_0$\_Vacc$_1$ and  -0.078 in No$_0$\_Conf-Vacc$_1$, which is due to the control efforts of the other group. For the controlling subpopulation the effectiveness  of their control effort has concomitantly weakened by  +.025, +.017 and +.047 on the group's infection load (w.r.t co-operation).
Third; we note that the interaction of the control strategies  has essentially weakened in the competitive scenarios.\footnote{In the co-operative scenarios with homogeneous spreading the interaction effect is the same as in Experiment 1.} We point out that the competitive configurations with No$_0$ in which the  counter strategies results in -.135 for Conf$_1$,  -.054  for Vacc$_{1}$ separately, and -.191 for Conf-Vacc$_1$ combined, which shows  an insignificant interaction gain of -.002 on the infection load.\\
Finally, we point out the competitive scenario with joint confinement strategy, Conf$_0$-Conf-Vacc$_1$ which shows the second best decrease of infection load outcome at -.220 for the confining group and -.254  for the confining-vaccinating group, and which represents the best in-group improvement of the infection load over all strategies.  The improved infection load for the out-group  w.r.t to the Nash joint confinement-vaccination strategy is related to the \emph{limited capacity} of the vaccine administration fixed in the model. In the competing scenario the fixed vaccine administration capacity (10 subjects per  lag of 6 time steps) is allocated to a single subpopulation which leads to the full number of individuals vaccinated per instance and a shorter time to cover the whole subpopulation. When both subpopulations share the limited capacity of vaccination administration, its implementation becomes less effective.  \\
% np.matrix([ [.543  , .469],   [ .593 , .593 ] , [.525 , .444 ] ]) -  .612
% matrix([[-0.069, -0.143],  [-0.019, -0.019],  [-0.087, -0.168]])
% competing confinement vs co-op - HOMOGENEOUS SPREADING
% np.matrix([.543  , .469]) -.453
% matrix([[0.09 , 0.016]])
A part of the control interaction effect and also the in-group effectiveness of control strategies should be related to the particular adjacencies of the near homogeneous mixing of the subpopulations. In the second instalment of Experiment 2, a block-localised spreading of the two subpopulations is used which can clarify the impact of subpopulation topology on the effectiveness of control scenarios.

\begin{table}[H] % changed h! to H
\centering
{
\renewcommand\arraystretch{1}

\begin{tabular}{|c*{4}{|c}|}
\multicolumn{3}{c}{} \\
\hline
& No$_1$ & Conf$_1$ & Vacc$_1$ & Conf-Vacc$_1$\\
\hline
No$_0$  & 0 & \backslashbox[]{-.008}{-.160} &\backslashbox[]{-.002}{-.068}&\backslashbox[]{-.009}{-.250}\\ 
\hline
Conf$_0$ & \diagbox[]{-.160}{-.008} & {-.160$^\star$} & \diagbox[]{-.162}{-.077 } &\diagbox[]{-.164}{-.255} \\
\hline
Vacc$_0$ & \diagbox[]{-.068}{-.002}  & \diagbox[]{-.077}{-.162}     &  -.036$^\star$    &  \diagbox[]{-.045}{-.231}\\
\hline
Conf-Vacc$_0$ & \diagbox[]{-.250}{-.009}   &     \diagbox[]{-.255}{-.164}           &   \diagbox[]{-.231}{-.045}    &   {\bf {-.238$^\star$}}\\
\hline
\end{tabular}
}
\vskip 3mm
\caption{The pay-off matrix, $P_{\textit{IL}}$, of the averaged {\bf infection load} change  - with respect to the infection
load .602 of the No$_0$-No$_1$ co-operative scenario, in competing two-group control scenarios with partially shuffled {\bf block-localised} group spreading in random social-contact graphs $\mathcal{G}_A(V,E_A)$, $n_{\rm iter} = 50$, SD $\approx$ 0.005 - 0.02). In the block-localised spreading configuration larger SD occur because initially infected nodes may be allocated unevenly over the groups. The network and epidemic parameters correspond to Tab. \ref{tab:infection_load}.}
   \label{tab:blocked_groups_infection_loads}
\end{table}
The observations in the competing control scenarios with the \emph{block-localised} spreading of the two subpopulations are again derived from the pay-off matrix of the group-specific infection load changes, Tab. \ref{tab:blocked_groups_infection_loads}. First we observe the optimal control solution  and the Nash equilibrium in the minimisation of the epidemic burden by the two subpopulations, which is again to deploy a co-operating  Conf-Vacc$_0$-Conf-Vacc$_1$ strategy by both subpopulations, leading to a gain of -.238 on the infection load. Further we observe that due to the partial geographical  separation, much less effect from the outside group's strategy is sustained, while the cross-interaction of the within-group strategy is more prominent again. A gain of -.160 on the infection load is reached by confinement, and -.068 is reached by vaccination, which shows  an interaction gain of -.022 in the infection load of -.250  when the strategies are combined. For the in-group without any control strategy all these out-group strategies have marginally no impact anymore on the in-group infection load. The latter is an immediate consequence of the partial geographic isolation of the two subpopulations.

\subsection{Control scenarios compounding the infection load reduction and the policy preference ranking \label{subsec:CS_compounded_objective} }

The deployment of epidemic control strategies is dependent on population socio-political factors in policy making \citep{MacIntyreEtAl2021,Reicher2020}. In our approach, the compounding method inserts a socio-political penalty to implementing a policy which is not preferred by the subpopulation, which alternatively can also be defined as a proxy cost in each subpopulation for deploying their respective control policies. The multiplicative weights method, Eqs. (\ref{eq:policyweights}), is used to assess two configurations, one in which the minimization of the infection load (IL) is ranked higher than the adherance to the control policy preference ranking (PR), with $O_{\textit{IL},\textit{PR}} = [1, 2]$ and the second configuration in which the minimization of the infection load and the adherance to the control policy preference ranking are equally valued $O_{\textit{IL},\textit{PR}} = [1,1]$.\footnote{To allow the comparison of the two configurations, the ranking intensity is kept fixed to $I=5$.}
The compounded pay-off matrix, $P_C$, for the weighted objectives can then be obtained as
\begin{eqnarray} 
P_C \ = \ \theta_1 P_{\textit{IL}} - \theta_2 P_{\textit{PR}},&\ \ \ & P_{\textit{PR}}\ = \ [w_0(I_0)_i\,, w_1(I_1)_j]  \label{eq:compounded_payoff_general}
\end{eqnarray}
where the infection load payoff matrix, $P_{\textit{IL}}$, is derived from Tab. \ref{tab:homogenous_groups_infection_loads} (or Tab \ref{tab:blocked_groups_infection_loads} for block-localisation), the weights, $\theta_i$, are calculated using Eq. (\ref{eq:policyweights}) with the rankings $O_{\textit{IL},\textit{PR}}$ for each configuration, while the weights $w_i$ follow from the rankings $O_{0/1}$ in Eq. (\ref{eq:policyrankings}) and constitute the entries of the policy ranking pay-off matrix $P_{\textit{PR}}$.\footnote{To balance the infection load factor (IL) with the control policy ranking (PR), the former has been normalised such that $\textit{IL}_{\textit{No}_0\_\textit{No}_1} = 1$, in Tabs \ref{tab:homogenous_groups_infection_loads} and \ref{tab:blocked_groups_infection_loads}.}\\
With different control policy preferences in both subpopulations, the Nash equilibrium can shift away from the co-operative Conf-Vacc$_0$-Conf-Vacc$_1$ strategy, Tabs \ref{tab:homogenous_groups_infection_loads}, \ref{tab:blocked_groups_infection_loads}. Depending on the policy ranking preferences ($O_0$ and $O_1$) and their intensities ($I_0$ and $I_1$) in each subpopulation,  the priority of the objectives of infection load reduction and policy preference or proxy cost ($O_{\textit{IL},\textit{PR}}$), and the type of population spread (homogeneous vs block-localised); the Nash Equilibrium will move across  the respective payoff matrices, $P_C$.  From the Nash equilibria, which are tabulated in Tabs \ref{tab:NE_homogeneous} and \ref{tab:NE_blocked}, one observes more variability in the case of a homogeneous spreading of subpopulations. This observation confirms the increased mutual effect of the competing control policies due to the higher number of between-group contacts in this type of spreading. In the homogeneous spreading configuration, the effect of an increased policy ranking intensity moves the Nash Equilibrium towards a competitive strategy in which the proper chosen strategy has a higher preference.
The latter effect is more outspoken when the objective of the infection load is weighted on a par with the policy preference ranking, $O_{\textit{IL},\textit{PR}} = [1, 1]$. In the socially more acceptable prioritisation of minimizing the infection load above adhering to policy ranking preference, $O_{\textit{IL},\textit{PR}} = [1, 2]$, the effect of partisanship is less diverse, but still results in a competitive Nash equilibrium  of Conf$_0$-Conf-Vacc$_1$ strategies (top versus bottom panels of Tabs \ref{tab:NE_homogeneous} and \ref{tab:NE_blocked}).\\

% \begin{table}[h!]
% \centering
% {
% \renewcommand\arraystretch{1}

% \begin{tabular}{|c*{4}{|c}|}
% \multicolumn{3}{c}{} \\
% \hline
% & No$_1$ & Conf$_1$ & Vacc$_1$ & Conf-Vacc$_1$\\
% \hline
% No$_0$          & \diagbox[]{.744}{.822} & \diagbox[]{.657}{.602} &\diagbox[]{.717}{.670} &\diagbox[]{.636}{.548} \\ 
% \hline
% Conf$_0$        & \diagbox[]{.602}{.735} & \diagbox[]{.568}{.568} & \diagbox[]{.541}{.577} &\diagbox[]{.485}{.460} \\
% \hline
% Vacc$_0$        & \diagbox[]{.737}{.795} & \diagbox[]{.644}{.541} & \diagbox[]{.760}{.694} &\diagbox[]{.672}{.556} \\
% \hline
% Conf-Vacc$_0$   & \diagbox[]{.559}{.714} & \diagbox[]{.471}{.485} & \diagbox[]{.567}{.605} & \diagbox[]{.493}{.482}\\
% \hline
% \end{tabular}
% }
% \vskip 3mm
% \caption{Payoff 8$\_$8$\_$8$\_$hom$\_$12}
%   \label{tab:Payoff_8_8_5_hom_12}
% \end{table}
% % group_0
% % 0,744	0,657	0,717	0,636
% % 0,602	0,568	0,541	0,485
% % 0,737	0,644	0,760	0,672
% % 0,559	0,471	0,567	0,493
% % Group_1
% % 0,822	0,602	0,670	0,548
% % 0,735	0,568	0,577	0,460
% % 0,795	0,541	0,694	0,556
% % 0,714	0,485	0,605	0,482

% Table generated by Excel2LaTeX from sheet 'NashTable_all'
\begin{table}[H] % changed h! to H
  \centering
\renewcommand\arraystretch{1.2}
\begin{tabular}{|c*{4}{|c}|}
\multicolumn{3}{c}{} \\
\hline
 \bf  hom \& [1,2] &   No$_1$ &    Conf$_1$ &    Vacc$_1$ &   Conf-Vacc$_1$ \\
\hline
      No$_0$ &   -    &   -    &    -   &  - \\
\hline
      Conf$_0$ &   -    &   -    &    -   & (4-7, 3-7) \\
\hline
      Vacc$_0$ &  -     &    -   &   -    & -  \\
\hline
      Conf\_Vacc$_0$ &   -    & \textcolor[rgb]{ .439,  .188,  .627}{} &    -   & (3, 3-7) \\
\hline  \hline
  \bf hom \& [1,1] &    No$_1$ &    Conf$_1$ &    Vacc$_1$ &    Conf-Vacc$_1$ \\
\hline
    No$_0$ &   -    &   -    &    -   & (7, 3) \\
\hline
   Conf$_0$ &   -    &   -    & (4-7, 4-7) & \textcolor[rgb]{ .439,  .188,  .627}{(3, 3-5)$^*$}, (4-6, 3) \\
\hline
   Vacc$_0$ &   -    & (3, 6-7) &    -   &  -  \\
\hline
   Conf\_Vacc$_0$ &    -   & \textcolor[rgb]{ .439,  .188,  .627}{(3, 3-5)$^*$} &   -    &  - \\
\hline
    \end{tabular}%
\vskip 3 mm 
      \caption{The Nash equilibria dependency on the intensity of the control policy ranking in two subpopulations, with {\bf homogeneous} subpopulation spreading, equal weighting  $O_{\textit{IL},\textit{PR}} = [1, 1]$  (bottom panel) or priority  weighting $O_{\textit{IL},\textit{PR}} = [1, 2]$ (top panel), of the infection load objective over the policy preference ranking. The brackets indicate the ranking intensities $(I_0,I_1)$ - or ranges - in both subpopulations, for which the Nash equilibrium occurs. (the \textcolor[rgb]{ .439,  .188,  .627}{$^*$} tag indicates the cases with two Nash equilibria) }
  \label{tab:NE_homogeneous}%
\end{table}%

% Table generated by Excel2LaTeX from sheet 'NashTable_all'
\begin{table}[H] % changed h! to H
  \centering
 \renewcommand\arraystretch{1.2}
\begin{tabular}{|c*{4}{|c}|}
\multicolumn{3}{c}{} \\
\hline
    \textbf{block \& [1,2]} & No$_1$ & Conf$_1$ & Vacc$_1$ & Conf-vacc$_1$ \\
\hline
    No$_0$ &   -    &     -  &    -   &  - \\
\hline
    Conf$_0$ &  -     &    -   &    -   & - \\
\hline
    Vacc$_0$ &   -    &    -   &    -   &  -\\
\hline
    Conf\_Vacc$_0$ &   -    & -  &    -   & (3-7, 3-7) \\
\hline \hline
    \textbf{block \& [1,1]} & No$_1$ & Conf$_1$ & Vacc$_1$ & Conf-vacc$_1$ \\
\hline
    No$_0$ &   -    &    -   &    -   &  - \\
\hline
    Conf$_0$ &   -    &   -    & (3-7, 5-7) & (3-7, 3-4) \\
\hline
    Vacc$_0$ &   -    &    -   &   -    &  - \\
\hline
    Conf-Vacc$_0$ &    -   & - &    -   &  - \\
\hline
    \end{tabular}%
    \vskip 3mm
      \caption{The Nash equilibria dependency on the intensity of the control policy ranking in two subpopulations, with {\bf block-localised} subpopulation spreading, equal weighting $O_{\textit{IL},\textit{PR}} = [1, 1]$ (bottom panel) or priority  weighting $O_{\textit{IL},\textit{PR}} = [1, 2]$ (top panel), of the infection load objective over the policy preference ranking. The brackets indicate the ranking intensities $(I_0,I_1)$ - or ranges - in both subpopulations, for which the Nash equilibrium occurs.}
  \label{tab:NE_blocked}%
\end{table}%

Finally, we note the special case  with an equal policy ranking intensity `3' in both subpopulations, with homogeneous spreading, with equal weighting of the infection load objective and policy preference ranking, which leads to a payoff matrix, Tab. \ref{tab:Payoff_3_3_5_hom_11}, with two Nash equilibria. In this particular configuration the two subpopulations can take recourse to either of the two competitive control scenarios Conf-Vacc$_0$-Conf$_1$ or Conf$_0$-Conf-Vacc$_1$. 

\begin{table}[H] % changed h! to H
\centering
{
\renewcommand\arraystretch{1}

\begin{tabular}{|c*{4}{|c}|}
\multicolumn{3}{c}{} \\
\hline
3$\_$3$\_$5$\_$hom$\_$11 & No$_1$ & Conf$_1$ & Vacc$_1$ & Conf-Vacc$_1$\\
\hline
No$_0$          & \diagbox[]{.301}{.434} & \diagbox[]{.248}{.250}           & \diagbox[]{.284}{.256} & \diagbox[]{.236}{.246} \\ 
\hline
Conf$_0$         & \diagbox[]{.250}{.381} & \diagbox[]{.229}{.229}          & \diagbox[]{.213}{.200} & \diagbox[]{\bf.179}{\bf.194} \\
\hline
Vacc$_0$         & \diagbox[]{.360}{.417} & \diagbox[]{.304}{.213}          & \diagbox[]{.374}{.270} & \diagbox[]{.321}{.251} \\
\hline
Conf-Vacc$_0$    & \diagbox[]{.275}{.369} & \diagbox[]{\bf.223}{\bf.179}    & \diagbox[]{.281}{.217} & \diagbox[]{.236}{.207} \\
\hline
\end{tabular}
}
\vskip 3mm
\caption{The compounded payoff matrix $P_{\textit{C}}$  (for configuration 3$\_$3$\_$5$\_$hom$\_$11) for an equal policy ranking intensity, $I_0 = I_1 =3$,  in both groups, with homogeneous spreading of the two subpopulations and with equal weighting of the infection load objective and policy preference ranking, $O_{\textit{IL},\textit{PR}} = [1, 1]$, exhibits two Nash equilibria (in boldface). }
  \label{tab:Payoff_3_3_5_hom_11}
\end{table}

\section{Discussion and conclusion}
A probabilistic state description of the individual nodes extends the approach of the SIR epidemic diffusion on graphs and allows a detailed assessment of the epidemic burden. Our study explored the outcomes of subpopulations using co-operative or competitive  epidemic control scenarios by means of a probabilistic SIR dynamics with temporary confinement and stochastic vaccination on  social-contact graphs. 
A basic comparison of co-operative vaccination and confinement scenarios was carried out to explore the  effects of these control strategies in relation to 
 their economic proxy cost from the total amount of administered vaccine doses, the reduction of the number of social contacts and the onset time of the confinement.  In our study we opted to first assess the epidemic impact by using the \emph{infection load} measure, which averages the maximal infection probability of all nodes over the full duration of the epidemic.
A multiple testing approach on designed random graphs indicates, first, the earliest onset time of interventions results in the lowest infection load on the social-contact graph, and second, the combination of the strategies of vaccination and confinement of the population in the social-contact graph results in the lowest infection load.
A clear interaction effect of control strategies is observed. The combined effect of vaccination and confinement surpasses the sum of both strategies separately.  Under competing strategies between subpopulations this interaction effect is mostly lost in the homogeneous mixing of the competing populations. When the groups are nearly block-separated, the interaction effect re-emerges.
The control scenarios for competing subpopulations show a Nash-equilibrium in the infection load pay-off matrix when they are co-operating in a combined vaccination-confinement strategy. The obtained infection-load pay-off matrices for two competing subpopulations do show effects that are specific to the type of implemented vaccination and confinement here. In particular, a competition for the limited vaccination capacity shows a diminished overall improvement in a block-separated subpopulation in comparison to a configuration with a homogeneously spread subpopulation.

A scalarization method for multi-objective optimization was used  to integrate the objective of the minimization of the infection load with a penalty - or proxy cost - for deploying a control policy with lower preference ranking.
The shifting of the Nash equilibrium over the compounded payoff matrix was observed for differing control policy preferences, and their intensities, in the two subpopulations.
In particular, the intensity of the control policy rankings and the prioritization  - or not - of the infection load reduction over policy ranking adherence, influence the proneness of the Nash equilibrium to shift. The  variability of the Nash equilibrium
was also shown to be higher in the homogeneous spreading of the  subpopulations in comparison to the block spreading. With homogeneous spreading, an increased policy ranking
intensity will move the Nash Equilibrium towards a competitive scenario which includes a more preferred strategy. This effect is more prominent when the policy preference ranking  objective is on par with the objective of the infection load reduction.

In our future work, a systematic exploration of cost factors (economic attrition and healthcare expenditure) due to confinement variations  and vaccination methods is envisaged (multiple types, repeated administration). Furthermore, the extension of the epidemic  SEIRSD model into a probabilistic epidemic diffusion model on social-contact networks will allow to cover seasonal diseases. Finally instead of randomly generated configurations, more realistic distributions of contact ego-networks, and  an improved allocation of individual infection and recovery rates can be implemented (according to age or medical preconditions  prevailing in subpopulations). Moreover, the effects can be studied when the control scenarios  and e.g. the intensities of the infectious exposures between nodes are modulated continuously over time.

\section*{Statements and Declarations}
The authors declare no competing interests.\\
The datasets generated during the current study are available in the OSF repository,  \href{https://osf.io/245ts/}{https://osf.io/245ts/}

\section*{Appendix: formal implementation of adjacency } 
\renewcommand{\theequation}{A.\arabic{equation}}% \thechapter.\arabic{equation}
The infection load of a SIR-epidemic on a graph depends on the structure of the social contact graph, epidemic infection and recovery rates of the population and spreading of the subpopulations, and specific implementation and parametrisation of the control scenarios. The specific implementations of the social-contact graph features for the reported simulations are provided in this appendix.
 A sorted adjacency matrix is obtained from random sampling of a row-sorted and column sorted random matrices. The randomised adjacency matrix is obtained by correlated row-column shuffling. The final probabilistic adjacency matrix is obtained by weighting the sorted and shuffled matrices with the \emph{skew} parameter. Clustering and cliques formation occurs in the dense ordered diagonal zone, hence \emph{skew} decreases clustering. \\
{\footnotesize
\hspace*{1em} Sorted: \hfill [a,b] = [1.1,0.9]\\
\hspace*{2em} random   $ \beta$(a,b,[N,N]), sort (axis=1), slice $\to$ sorted$\_$triu$\_$H, symmetrize $\to$  A$\_$sorted$\_$H \\
\hspace*{2em} random   $ \beta$(a,b,[N,N]), sort (axis=0), slice $\to$ sorted$\_$triu$\_$V, symmetrize $\to$  A$\_$sorted$\_$V \\
\hspace*{2em} sample by random$\_$idx \{A$\_$sorted$\_$H, A$\_$sorted$\_$V\} $\to$ A$\_$sorted\\
\hspace*{1em} Shuffled: \\
\hspace*{2em} shuffle (axis=1) sorted$\_$triu$\_$H, move 0's left, slice symmetric matrix $\to$ A$\_$shuffled$\_$H\\
\hspace*{2em} shuffle (axis=0) sorted$\_$triu$\_$V, move 0's down, slice symmetric matrix $\to$ A$\_$shuffled$\_$V\\
\hspace*{2em} sample by same random$\_$idx \{A$\_$shuffled$\_$H, A$\_$shuffled$\_$V\} $\to$ A$\_$shuffled\\
\hspace*{1em} Weighted: \hfill skew = 0.5 
\begin{eqnarray}
 \hspace*{2em} {\rm A}\_{\rm fin}  = {\rm A}\_{\rm sorted} \times (1 - {\rm skew}) + A\_{\rm shuffled} \times {\rm skew} \label{eq:skew}
\end{eqnarray}
\hspace*{1em} Binary:\hfill N$\_$connect = 50, N$\_$confined$\_$A = 20 \\
\hspace*{2em} free : N$\_$ = N$\_$connect,  confined : N$\_$ = N$\_$confined$\_$A\\
\hspace*{2em} A = (A$\_$fin $> =$ (N - N$\_$)/N) \\
% }\\
% Masking of adjacency is applied for competing group confinement strategies. In particular the between group adjacency is implemented using group averaged connectivity parameters.\\
% {\footnotesize 
Competing confinement:\\
% \hspace*{1em}    confined groups  0 and 1  $\to$  N$\_$01 = (N$\_$confinement$\_$A0 + N$\_$confinement$\_$A1)/2\\
\hspace*{1em}    confined groups   0  xor 1 :
\begin{eqnarray}
{\rm N}\_{01} &=& ({\rm N}\_{\rm confinement}\_{\rm A0} + {\rm N}\_{\rm connect}\_{\rm A})/2 \label{eq:mixed_threshold}
\end{eqnarray}
%#\hspace*{1em}    confined groups   1  $\to$    N$\_$01 = (N$\_$connect$\_$A + N$\_$confinement$\_$A1)/2\\
%\hspace*{1em}    A$\_$sub00 = ((Mask$\_$00*A$\_$fin)>=(N-N$\_$00)/N )$\times$ 1 \\
\hspace*{1em}    A$\_$sub01 = ( Mask$\_$01 $\times$ A$\_$fin $>=$ (N-N$\_$01)/N )  \\
%\hspace*{1em}    A$\_$sub11 = ((Mask$\_$11*A$\_$fin)>=(N-N$\_$11)/N )$\times$ 1 \\
\hspace*{1em}    A$\_$confined$\_$groups = A$\_$sub00 + A$\_$sub01 + A$\_$sub11
}

\bibliography{SIR_on_Graph_dynamics.bib}
\end{document}